\documentclass{easychair}

\usepackage{makeidx}
\usepackage{multirow}

\lstdefinestyle{codeStyle}
{
        language=Java,
        frame=single,  
        basicstyle=\footnotesize,
        captionpos=b,
        showstringspaces=false,
        showspaces=false,
        extendedchars=true,
        linewidth=1\linewidth,
        breaklines=true,
        float=phtb  
}

\newcommand{\xf}[1]{Figure~\ref{#1}}

\newcommand{\xs}[1]{Section~\ref{#1}}

\newcommand{\xt}[1]{Table~\ref{#1}}
\newcommand{\xl}[1]{Listing~\ref{#1}}

\newcommand{\java}{{Java\index{Java}}}

\newcommand{\todo}[0]
{
	{\Large \[TODO\]}
}

\newcommand{\file}[1]{\url{#1}\index{Files!#1}}
\newcommand{\tool}[1]{\texttt{#1}\index{Tools!#1}}

\newcommand{\api}[1]{\texttt{#1}\index{API!#1}}

\newcommand{\marf}[0]{MARF\index{MARF}\index{Frameworks!MARF}\index{Libraries!MARF}}
\newcommand{\dmarf}[0]{DMARF\index{MARF!Distributed}\index{Frameworks!Distributed MARF}\index{Libraries!Distributed MARF}}

\newcommand{\lucidL}[1]{{$\mathit{Lucid}$}($L$) }

		{}

\def\myvert{\raise 2.27pt \hbox{\vrule depth 0pt height 8pt width 0.2mm}}
\def\myarrow{\hspace*{0.43mm}%
             \raise 2.29pt\hbox{\vrule depth 0pt height 8pt width 0.16mm}%
             \hspace*{-0.32mm}%
             $\longrightarrow$
             \ %
             }

\newcommand{\cmdline}[1]{\texttt{#1}\index{Commands!#1}\index{#1}}

\makeindex

\newcommand{\lablinuxdistro}{Scientific Linux 6.6\index{Linux!Scientific Linux 6.6}}

\newcommand{\openesbdistro}{OpenESB 2.31\index{OpenESB!OpenESB 2.31}\index{NetBeans!NetBeans 7.1}}
\newcommand{\netbeans}{NetBeans\index{NetBeans}}
\newcommand{\tomcatdistro}{Tomcat 6\index{Tomcat!Tomcat 6}}
\newcommand{\tomcat}{Tomcat\index{Tomcat}}
\newcommand{\glassfishdistro}{GlassFish 2.x/3.0.1\index{GlassFish!GlassFish 2.x/3.0.1}}
\newcommand{\glassfish}{GlassFish\index{GlassFish}}

\newcommand{\course}{SOEN487}

\newcommand{\coursewww}{course web page}

\begin{document}

\title{Service-Oriented Architectures and Web Services:\\Course Tutorial and Lab Notes}
\titlerunning {Service-Oriented Architectures and Web Services: Course Tutorial and Lab Notes}

\author
{
	Serguei A. Mokhov
	\hspace{1cm} Shahriar Rostami
	\hspace{1cm} Hammad ALi
	\hspace{1cm} Yuhong Yan\\
	Computer Science and Software Engineering\\
	Faculty of Engineering and Computer Science\\
	Concordia University\\
	Contact: \url{{mokhov,yuhong}@cse.concordia.ca}
}
\authorrunning{Mokhov \emph{et al}.}

\maketitle

\begin{abstract}
This document presents a number of quick-step instructions to get started on writing mini-service-oriented
web services-based applications using
{\openesbdistro},
{\tomcatdistro},
{\glassfishdistro} with BPEL support,
and {\java} 1.6+ primarily
in {\lablinuxdistro} with user quota restrictions. While the tutorial notes are oriented towards the
students taking the {\course} on service-oriented architectures (SOA) at Computer Science
and Software Engineering (CSE) Department, Faculty of Engineering and Computer Science (ENCS),
other may find some of it useful as well outside of CSE or Concordia. The notes are compiled
mostly based on the students' needs and feedback.
\end{abstract}

\tableofcontents
\listoffigures
\lstlistoflistings
\listoftables
\clearpage


\section{Introduction}
\label{sect:introduction}

\noindent
{\bf NOTE:} these notes are undergoing an update for the current term.

\noindent
{\bf NOTE:} A copy these notes may also be released more frequently on {\coursewww},
SourceForge while pending approval of updates on arXiv. Thus, available ``mirrors''
for these are:

\begin{itemize}
\item
{\coursewww} (most up-to-date)
\item 
\url{https://sourceforge.net/projects/atsm/files/SOEN%20WS/Winter%202015/}
\item
\url{http://arxiv.org/abs/0907.2974}
\end{itemize}

\section{Tentative Lab and Tutorial Schedule}
\label{sect:tut-lab-schedule}

In the tutorials the tutor demonstrates the environment, tools,
etc. (1 hour). In the lab, the students reproduce the tutorial
material with the assistance of the lab instructor. The hands-on
aspect is primarily in the lab. Including assistance with the
project and assignments.

\begin{description}
\item [Week 2]
	Lab environment,
	see \xs{sect:lab-env}.

\item [Week 3]
	Explicit XML parsing using DOM, xpath, JAXB as well as HTTP processing with {\java}; see references
	in \xs{sect:prerequisites}.

\item [Weeks 4--5]
	SOAP services,
	see \xs{sect:preparatory-notes}.
	Project support.	

\item [Weeks 6--7]
	REST services,
	see \xs{sect:restful-ws}.
	Project support.	

\item [Weeks 9--11]
	BPEL,
	see \xs{sect:bpel}.
	Project support.	

\item [Weeks 12--15]
	Project support.

\end{description}

\subsection{Prerequisites}
\label{sect:prerequisites}

Basic knowledge and prerequisite skills required to grasp in order
to do the introductory assignments and have a deeper knowledge of
the subject include:

\begin{enumerate}

\item
	Understanding the concepts behind web services (WS)\index{web services} in the context of distributed
	system and distributed computing. How WS came to be and what are its major advantages and
	disadvantages.

	Introductory notes on WS can be found in the NetBeans and Oracle tutorials
	\cite{netbeans-kb-ws-jax-ws,java-webservices}.

\item
	Understanding HTTP POST and GET in general is important.
	More specifically programmability of HTTP POST and GET with {\java},
	as e.g. given in \cite{java-http-post-get-requests}.

\item
	Manual XML parsing with {\java} is important to understand
	especially when dealing with parsing a custom application's
	XML files to extract or represent the data.
  
	A simple example is available at \cite{java-xml-parsing-dom-stringreader},
	and more realistic application-specific examples in \cite{marf}'s
	\api{NeuralNetwork} implementation and its testing application \api{TestNN} \cite{marf-test-nn}:

	\url{http://marf.cvs.sf.net/viewvc/marf/apps/TestNN/}

	\url{http://marf.cvs.sf.net/viewvc/marf/marf/src/marf/Classification/NeuralNetwork/}

\item JAXB and Java \cite{netbeans-kb-ws-jaxb}

\item Programming in {\java} \cite{javanuttshell}, JSP \cite{jsp}, Servlets \cite{servlets}, JavaScript
\end{enumerate}

\subsection{Related Work and Reading}
\label{sect:related-work}

This section compiles a set of related work and reading references.

\begin{enumerate}
	\item
Reference texts:

SOA and WS:
\cite{ibm-redbook-patterns-soa-ws-2004,%
soa-ws-data-management-2008,%
soa-semantics-processes-agents-2005}

JavaScript, AJAX, WWW:
\cite{programming-www-2010}

	\item
Why not CORBA?

\cite{rise-fall-corba-2006}

	\item
BPEL, WSC problem modeling and overview, algorithms, SOAP routing, and AI planning:

\cite{repairing-service-composition-2010,%
compat-reparation-ws-processes-2010,%
ws-enabled-lab-2009,%
koenig-ws-bpel-2007,%
wsc08,%
ws-soap-routing-2008,%
wsben-2009,%
wsc-ai-planning-survey,%
comparative-ill-ai-wsc-2005,%
wspr-scalable-wsc-algo-2007,%
wsc-framework-int-programming-2008,%
planning-monitoring-wsc-2004,%
automated-planning-2004,%
type-aware-wsc-bool-sat-2008,%
auto-wsc-and-or-graph-2008,%
plan-graph-algo-semantic-wsc-2008,%
standards-based-service-repo-2008}

	\item
Formal methods in WS:

\cite{dl-fm-wsp-modeling-2008,%
model-faults-in-ws-processes-2009,%
service-interaction-2009,%
fm-orchestration-2004,%
reasoning-ws-process-algebra-2004,%
fm-semantics-ctrl-flow-ws-bpel-2005,%
fm-verification-bpel4ws-2004,%
service-science-and-soss-2008}

	\item
Applications:

\cite{ws-mash-up-home-library-2008,%
ws-sim-map-forest-fire-2008,%
wsc-gis-2008,%
dmarf-web-services-cisse08}

	\item
Lecture notes and slides:

\cite{soen-691a-487-winter2011,%
soen-691a-487-winter2011-intro,%
soen-691a-487-winter2011-service,%
soen-691a-487-winter2011-distrib-computing,%
soen-691a-487-winter2011-xml,%
soen-691a-487-winter2011-soap,%
soen-691a-487-winter2011-wsdl,%
soen-691a-487-winter2011-ws-programming,%
soen-691a-487-winter2011-rest,%
soen-691a-487-winter2011-uddi,%
soen-691a-487-winter2011-bpel%
}

\end{enumerate}

\section{Preparatory Notes on Web Services}
\label{sect:preparatory-notes}

\subsection{Create a Web Service with {\netbeans}, test Web Service, Observe SOAP Messages}

Reading and practice:

\begin{itemize}
	\item 
Build: \api{ManufacturerService.processPO} as Exercise 1 as illustrated in \cite{netbeans-kb-ws-jax-ws}.

	\item 
Knowledge: Chapter 1 in \cite{java-webservices}.

	\item 
One can create a Web Service as {\em Java Web Application} or as {\em Maven Web Application}.
Maven \cite{maven} is a project management tool on top of Ant \cite{ant}.

	\item 
One needs to compile and deploy the service before you testing it.
\end{itemize}

\subsection{Consume Web Service with Java classes, JSP and Servlet}

Use the referenced tutorial \cite{netbeans-kb-ws-jax-ws}
as well as see the login example in \xs{sect:simple-ws-app-testing}.

\subsection{Complex Data Types Used in Web Services, JAXB Binding}

Consult the two tables (\xt{tab:jaxb-mapping-xml-types}, \xt{tab:jaxb-mapping-xml-types-to-java})
are from \cite{java-webservices} as a reference,
more specifically from \cite{java-webservices-jaxb}.

\begin{table}[htpb]%
\centering
\caption{JAXB Mapping of XML Schema Built-in Data Types \cite{java-webservices-jaxb}}
\label{tab:jaxb-mapping-xml-types}
\begin{tabular}{|l|l|}\hline
XML Schema Type         & Java Data Type\\\hline\hline
\api{xsd:string}        & \api{java.lang.String} \\\hline
\api{xsd:integer}       & \api{java.math.BigInteger} \\\hline
\api{xsd:int}           & \api{int} \\\hline
\api{xsd.long}          & \api{long} \\\hline
\api{xsd:short}         & \api{short} \\\hline
\api{xsd:decimal}       & \api{java.math.BigDecimal} \\\hline
\api{xsd:float}         & \api{float} \\\hline
\api{xsd:double}        & \api{double} \\\hline
\api{xsd:boolean}       & \api{boolean} \\\hline
\api{xsd:byte}          & \api{byte} \\\hline
\api{xsd:QName}         & \api{javax.xml.namespace.QName} \\\hline
\api{xsd:dateTime}      & \api{javax.xml.datatype.XMLGregorianCalendar} \\\hline
\api{xsd:base64Binary}  & \api{byte[]} \\\hline
\api{xsd:hexBinary}     & \api{byte[]} \\\hline
\api{xsd:unsignedInt}   & \api{long} \\\hline
\api{xsd:unsignedShort} & \api{int} \\\hline
\api{xsd:unsignedByte}  & \api{short} \\\hline
\api{xsd:time}          & \api{javax.xml.datatype.XMLGregorianCalendar} \\\hline
\api{xsd:date}          & \api{javax.xml.datatype.XMLGregorianCalendar} \\\hline
\api{xsd:g}             & \api{javax.xml.datatype.XMLGregorianCalendar} \\\hline
\api{xsd:anySimpleType} & \api{java.lang.Object} \\\hline
\api{xsd:anySimpleType} & \api{java.lang.String} \\\hline
\api{xsd:duration}      & \api{javax.xml.datatype.Duration} \\\hline
\api{xsd:NOTATION}      & \api{javax.xml.namespace.QName} \\\hline
\end{tabular}
\end{table}

\paragraph{\api{JAXBElement}.}

When XML element information can not be inferred by the derived Java representation
of the XML content, a \api{JAXBElement} object is provided. This object has methods
for getting and setting the object name and object value.

\paragraph{Java-to-Schema.}

The referenced \xt{tab:jaxb-mapping-xml-types-to-java} shows
the default mapping of Java classes to XML data types.

\begin{table}[htpb]%
\centering
\caption{JAXB Mapping of XML Data Types to Java Classes \cite{java-webservices-jaxb}}
\label{tab:jaxb-mapping-xml-types-to-java}
\begin{tabular}{|l|l|}\hline
Java Class                                    & XML Data Type \\\hline\hline
\api{java.lang.String}                        & \api{xs:string} \\\hline
\api{java.math.BigInteger}                    & \api{xs:integer} \\\hline
\api{java.math.BigDecimal}                    & \api{xs:decimal} \\\hline
\api{java.util.Calendar}                      & \api{xs:dateTime} \\\hline
\api{java.util.Date}                          & \api{xs:dateTime} \\\hline
\api{javax.xml.namespace.QName}               & \api{xs:QName} \\\hline
\api{java.net.URI}                            & \api{xs:string} \\\hline
\api{javax.xml.datatype.XMLGregorianCalendar} & \api{xs:anySimpleType} \\\hline
\api{javax.xml.datatype.Duration}             & \api{xs:duration} \\\hline
\api{java.lang.Object}                        & \api{xs:anyType} \\\hline
\api{java.awt.Image}                          & \api{xs:base64Binary} \\\hline
\api{javax.activation.DataHandler}            & \api{xs:base64Binary} \\\hline
\api{javax.xml.transform.Source}              & \api{xs:base64Binary} \\\hline
\api{java.util.UUID}                          & \api{xs:string} \\\hline
\end{tabular}
\end{table}

\begin{enumerate}
	\item 
Binding WSDL with JAXB \cite{netbeans-kb-ws-jaxb}:

\begin{itemize}
	\item 
The class of a complex type should have an non-argument default constructor.
	\item 
The WSDL contains a \api{schemaLocation} to indicate an xml that describes the complex data type.
\end{itemize}

	\item 
Given a WSDL file, ask for implementing such a Web service:

\begin{itemize}
	\item 
Method 1: at the ``Web Service'' folder, right click, select ``Web service from WSDL''.
This creates a JAX-WS service.
Or you can find ``Web service from WSDL'' when you right click the project node.

	\item 
Method 2:
(1) first create an xml (i.e. the WSDL) to Java binding.
(2) create an empty Web Service.
(3) add operations described in the WSDL to the Web service, with the required inputs and outputs.
If the inputs and the outputs are complex types, JAXB binding is useful.
One needs to use the classes generated by JAXB binding (step 1).
If the service has only simple types, you do not need step (1).
Example: \api{CreditReportService} \cite{netbeans-kb-ws-jaxb}.
\end{itemize}
\end{enumerate}

\paragraph{Exercises for key points:}

\begin{enumerate}
	\item 
 Create a WS with simple and complex XML data types
	\item 
 Develop a SOAP WS with top-down and bottom-up approaches 
\end{enumerate}

\noindent
Related tutorials:

\begin{enumerate}
	\item 
 Top-down and bottom-up approaches tutorial e.g. in \cite{netbeans-kb-gs-axis}

	\item 
 Passing binary data in SOAP: a five-part tutorial in \cite{netbeans-kb-trails-web}
 
	\item 
 Develop Web service with complex XML data type with JAXB binding \cite{netbeans-kb-ws-jaxb}
\end{enumerate}

\section{RESTful Web Services}
\label{sect:restful-ws}

{\em REpresentational State Transfer (REST)}.
Follow {\netbeans} tutorial \cite{netbeans-kb-ws-rest}.
The lecture notes are in \cite{soen-691a-487-winter2011-rest}.
Delicious restful service demo at
\url{http://jmvidal.cse.sc.edu/talks/rest/delicious.html}:
use JavaScript to call a RESTful service and use JavaScript process
the results in JSON ({\em(JavaScript Object Notation)}
is a lightweight data-interchange format, \url{www.json.org}).

\includegraphics[width=\textwidth,page=1]{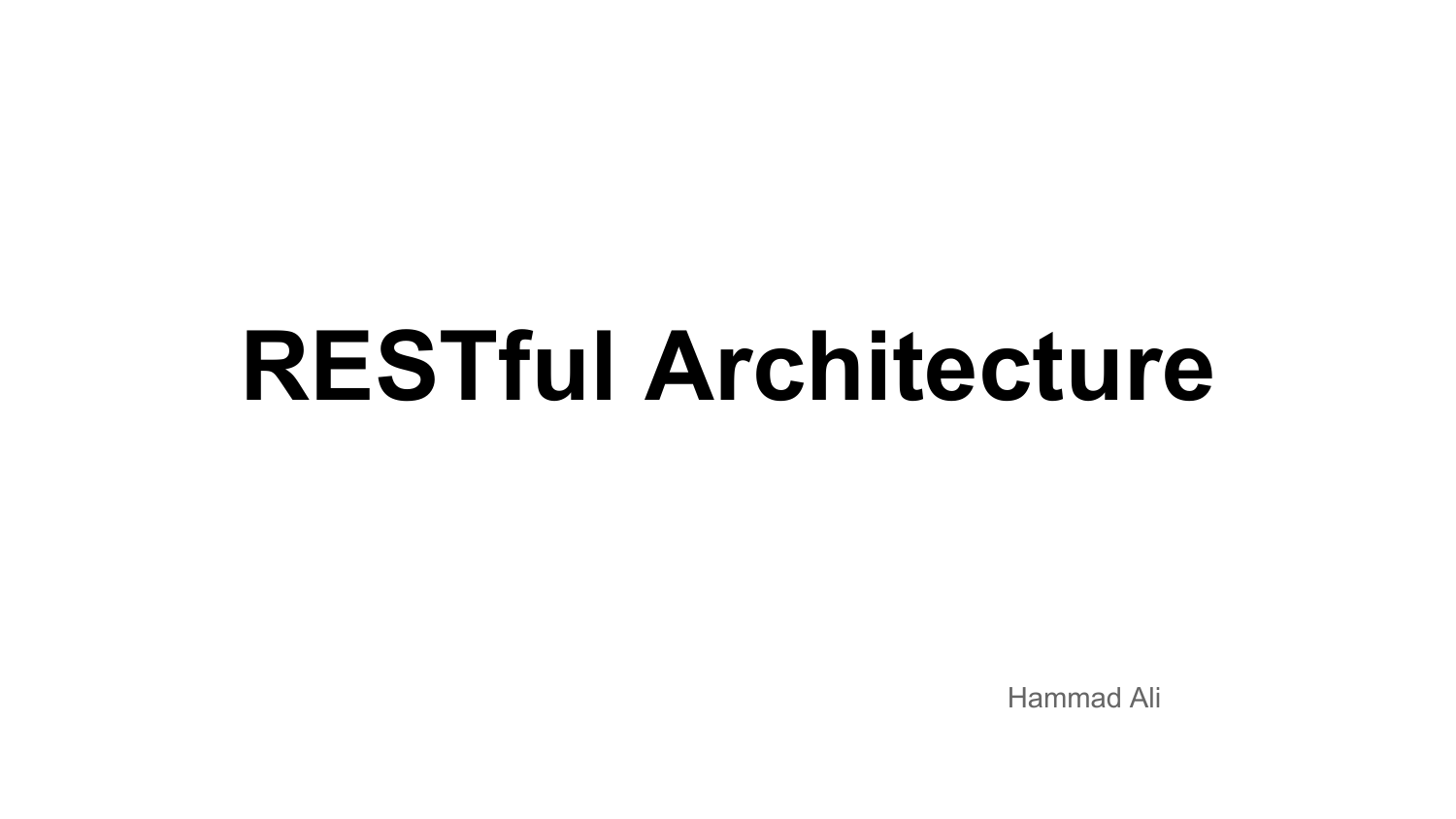}
\includegraphics[width=\textwidth,page=2]{images/restful-services.pdf}
\includegraphics[width=\textwidth,page=3]{images/restful-services.pdf}
\includegraphics[width=\textwidth,page=4]{images/restful-services.pdf}
\includegraphics[width=\textwidth,page=5]{images/restful-services.pdf}
\includegraphics[width=\textwidth,page=6]{images/restful-services.pdf}

\subsection{Exercises}

\subsubsection{HelloWorld}

Test the sample REST service \api{HelloWorld}, where
one can set and get the resource value in the resource \api{HelloWorld}.
Use HTTP GET/PUT to get and change the resource values.
Resource end point:

\url{http://localhost:8080/HelloWorld/resources/helloWorld}

%
%

\subsubsection{CustomerDB}

\api{CustomerDB} is a more elaborate example. We'd
use Derby, the default NetBeans's internal database
engine.

\subsection{Optional Exercise}

What would be required to convert your Exercise 4 from Assignment 1
to become RESTful services? Try doing it.

\section{BPEL}
\label{sect:bpel}

Reading and resources: \cite{ws-bpel-11,%
ws-bpel-20,%
koenig-ws-bpel-2007,%
bpelse,%
wiki:bpel,%
soen-691a-487-winter2011-bpel,%
fm-semantics-ctrl-flow-ws-bpel-2005,%
fm-verification-bpel4ws-2004}

\clearpage
\subsection{OpenESB 2.3.1}

\includegraphics[width=\textwidth,page=1]{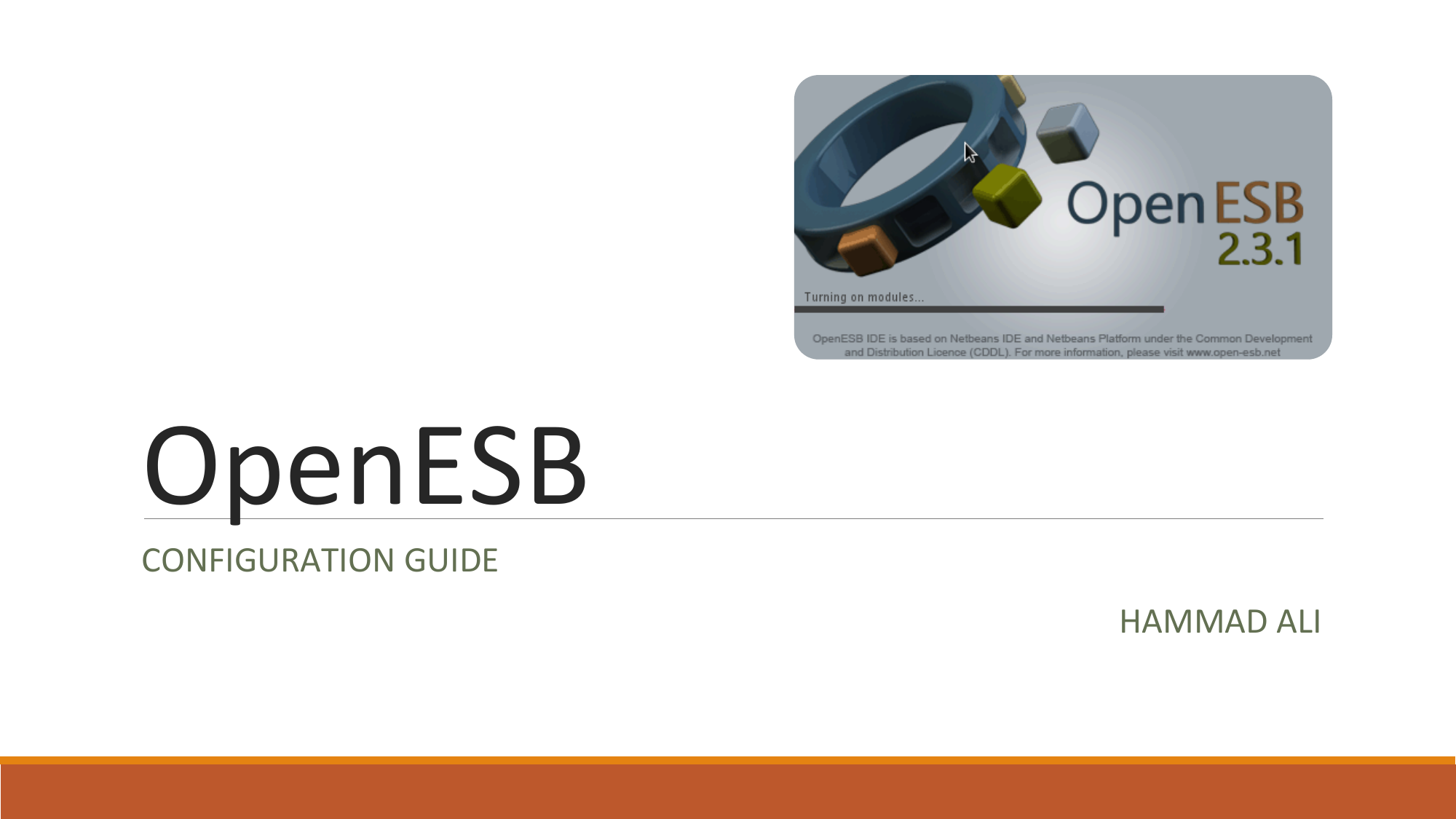}
\includegraphics[width=\textwidth,page=2]{images/openesb-glassfish.pdf}
\includegraphics[width=\textwidth,page=3]{images/openesb-glassfish.pdf}
\includegraphics[width=\textwidth,page=4]{images/openesb-glassfish.pdf}
\includegraphics[width=\textwidth,page=5]{images/openesb-glassfish.pdf}
\includegraphics[width=\textwidth,page=6]{images/openesb-glassfish.pdf}
\includegraphics[width=\textwidth,page=7]{images/openesb-glassfish.pdf}
\includegraphics[width=\textwidth,page=8]{images/openesb-glassfish.pdf}
\includegraphics[width=\textwidth,page=9]{images/openesb-glassfish.pdf}
\includegraphics[width=\textwidth,page=10]{images/openesb-glassfish.pdf}
\includegraphics[width=\textwidth,page=11]{images/openesb-glassfish.pdf}
\includegraphics[width=\textwidth,page=12]{images/openesb-glassfish.pdf}
\includegraphics[width=\textwidth,page=13]{images/openesb-glassfish.pdf}
\includegraphics[width=\textwidth,page=14]{images/openesb-glassfish.pdf}
\includegraphics[width=\textwidth,page=15]{images/openesb-glassfish.pdf}
\includegraphics[width=\textwidth,page=16]{images/openesb-glassfish.pdf}

\subsection{NetBeans 6.9.1 and Eclipse}

There is some chance BPEL would be available for the newer
platforms, but we are not holding our breath for it presently,
so for now we'll stick with the legacy setup in \xs{sect:bpel-legacy-setup}.

\subsection{Legacy}
\label{sect:bpel-legacy-setup}

BPEL plugins (GUI and the service engine) for now are known to work
``out-of-the-box''
with NetBeans 6.5.1 \cite{netbeans-651} (labs) and 6.7.1 \cite{netbeans-671} (download).
See notes about Linux commands and project directories in \xs{sect:linux-env-setup}.

Tools, resources and their locations:

\begin{itemize}
	\item NetBeans 6.5.1 \cite{netbeans-651} path in Linux in labs
	
	32-bit systems:
	
	\file{/encs/pkg/netbeans-6.5.1/root/bin/netbeans}
	
	Remote login to \texttt{computation.encs.concordia.ca} (64-bit):
	
	\file{/encs/ArchDep/i686.linux26-RHEL5/pkg/netbeans-6.5.1/root/bin/netbeans}
	
	\item NetBeans 6.5.1 download:
	
	\url{http://netbeans.org/downloads/6.5.1/index.html}

	\item NetBeans 6.7.1 download:
	
	\url{http://netbeans.org/downloads/6.7.1/index.html}
	
	\item BPEL Guide and Tutorials:
	
	\url{http://netbeans.org/projects/usersguide/downloads/download/NB61-SOAdocs.zip}

	\url{http://www.youtube.com/watch?v=a76RxkzB4Bg}

	\item BPEL Lecture Notes \cite{soen-691a-487-winter2011-bpel}
\end{itemize}

\subsubsection{Configuring NetBeans and GlassFish for BPEL}

The \texttt{ALL} option typically installs GlassFish 2.1~\cite{glassfish}
as well as Tomcat 6 bundled by default with NetBeans, as
well as some of the components. This includes some
of the BPEL~\cite{wiki:bpel} components as well. To complete all the needed
extensions for BPEL for GlassFish you'd need to download WSDL
extensions and Saxon shared libraries and deploy them within
your running GlassFish instance.
Download libraries for BPEL SE~\cite{bpelse},
specifically: \file{wsdlextlib.jar} and \file{saxonlib.jar};
these should go under ``Shared Libraries''.
That's all you need for your setup in the lab. For your
home computer you may need to download and install the
actual BPEL service engine component from the same
web page~\cite{bpelse}, called \file{bpelserviceengine.jar}, which
should go under ``Service Engines'' and NOT ``Shared Libraries''.

\subsubsection{BPEL Composite Applications}

GlassFish 2.1 is needed for legacy BPEL. 
E.g. see the tutorial from NetBeans referenced above.


Similarly, there are good application samples available
in the betbeans to start the process of a BPEL composite
application:
``New'' $\rightarrow$ ``Samples'' $\rightarrow$ ``SOA'';
specifically ``Travel Resevation Service'' and ``BPEL BluePrint 1''.

{\bf NOTE:} The referenced tutorial works best with NetBeans 6.5.1
and was found to have difficulties in NetBeans 6.7 (e.g. empty
\file{Output.xml} file is not being prompted for or produced). If you insist
on using NetBeans 6.7 be extra careful to the warning notes in
the tutorial web page.



\section{WS Reliability and Security}
\label{sect:ws-reliability-security}

{\todo}

\section{Packaging and Deployment}
\label{sect:packaging-and-deployment}
\label{sect:deployment}

%
%
%

\begin{itemize}
\item NetBeans
\item Ant \cite{ant}
\item Maven \cite{maven}
\end{itemize}

{\todo}

\section{{\glassfish}}
\label{sect:glassfish}

\subsection{Starting {\glassfish} as a Standalone Service}
\label{sect:glassfish-standalone}

This is to startup {\glassfish} outside of the {\netbeans}
environment as a standalone service for application deployment:

\url{http://download.oracle.com/docs/cd/E19798-01/821-1757/gglog/index.html}

\subsubsection{Windows}

{\glassfish} and configuration

\begin{enumerate}
	\item 
{\glassfish} installation directory: \verb+C:\Program Files\glassish-3.0.1+. Use
\api{\%GF\%} to represent this path.

You may need to make \api{\%GF\%} editable (change security) in order to be able to
log events, run modules, etc.

	\item 
{\netbeans} runs {\glassfish} from the specified domain, e.g.\\
\verb+C:\\users\\USERNAME\\.netbeans\\6.9\\config\\GF3\\domain1+.
Use \api{\%NG\%} to represent this path.

	\item 
To start {\glassfish} from \api{\%GF\%} (not inside {\netbeans}), run from command line:

\texttt{asadmin start-domain --verbose}
\end{enumerate}

\section{Marking Schema for the Assignments and Project}
\label{sect:marking-env}

Please refer to the project document.

\section{Lab Environment}
\label{sect:lab-env}

\subsection{Windows}
\label{sect:windows}

On ENCS Windows the software was not made
readily available (in particular more recent
NetBeans with the \texttt{ALL} option).

\subsection{Linux}
\label{sect:linux}

We are using {\lablinuxdistro} during the labs.
For your own work you can use any platform of
your choice, e.g. Windows or MacOS X on your
laptops. You will have to do the installation
and configuration of {\netbeans}, {\java}, {\glassfish},
or {\tomcat} and so on there yourself.

\subsubsection{Accounts}
\label{sect:accounts}

\newcommand{\mygrpshort}{kos691a2}
\newcommand{\mygrp}{\url{ko_soen691a_2}}
\newcommand{\mygrpfull}{\url{/groups/k/ko_soen691a_2}}

Under UNIX, disk space (for a sample account \texttt{\mygrpshort}) would be
accessible under e.g. {\mygrpfull}.
Under Windows (in case you need to access files from that OS), that path would be
\verb+\\filer-groups\v_groups\groups_unix\k\ko_soen691a_2+ (the ``S:'' drive).
There is a 1GB quota space available there and your in-school work related
to the assignments and courses can be put there, as the generated
data files can be large at times.

To figure out what is your group account, type the \cmdline{id} command
in a terminal, e.g.:

\begin{verbatim}
addams.mokhov [ko_soen691a_2] % id
uid=X(mokhov) gid=X(mokhov) groups=...,8896(kos691a2),...,X(mokhov)
\end{verbatim}

\subsubsection{Java 1.6}
\label{sect:java-16}

Java 1.6 is not a default {\java} in ENCS at this moment. You need to make
it default. In order to use this version all you need to do is
prepend:

\file{/encs/pkg/jdk-6/root/bin}

\noindent
to your \api{PATH} (the environment variable). To do so there are simple instructions:

\noindent
People using \tool{tcsh} (the default):

\begin{verbatim}
addams.mokhov [~] % setenv PATH /encs/pkg/jdk-6/root/bin:$PATH
addams.mokhov [~] % rehash
addams.mokhov [~] % java -version
java version "1.6.0_29"
Java(TM) SE Runtime Environment (build 1.6.0_29-b11)
Java HotSpot(TM) Server VM (build 20.4-b02, mixed mode)
\end{verbatim}

\noindent
People using \tool{bash}:

\begin{verbatim}
bash-2.05b$ export PATH=/encs/pkg/jdk-6/root/bin:$PATH
bash-2.05b$ java -version
java version "1.6.0_29"
Java(TM) SE Runtime Environment (build 1.6.0_29-b11)
Java HotSpot(TM) Server VM (build 20.4-b02, mixed mode)
\end{verbatim}

You can avoid typing the above commands to set the \api{PATH} each time you open a terminal
under Linux by recording it in \file{~/.cshrc}. If you
do not have this file in your home directory you can create one
with the following content (e.g. using \tool{vim}~\cite{vim}):

\begin{verbatim}
set path=( /encs/pkg/jdk-6/root/bin $path )
\end{verbatim}

\noindent
or copy an example from~\cite{mokhov-dot-cshrc-example}
and update the \api{path} to include the above directory to be the first on the list.
Thus, next time when one logs in and opens a terminal, Java 1.6
will always be the default. The same applies to the {\java} used when one clicks
on the NetBeans or Eclipse shortcuts in the graphical menu.

\subsubsection{NetBeans}
\label{sect:netbeans}

NetBeans~\cite{netbeans} is a major IDE of the supported for the course.

\paragraph{NetBeans 6.9.1}

NetBeans 6.9.1~\cite{netbeans-691} is accessible as a simple command \tool{netbeans}
or from the ``Applications'' $\rightarrow$ ``Programming'' $\rightarrow$ ``NetBeans''
menu with a corresponding icon. This version of NetBeans does not have support plug-ins
for SOA and BPEL at the time of this writing.

\paragraph{NetBeans 6.5.1}

This material is for legacy versions of NetBeans 6.5.1 and GlassFish 2 under
Scientific Linux 5.6 environment that support SOA, BPEL Designer and a run-time
service engine.
The legacy version NetBeans 6.5.1 \cite{netbeans-651} is accessible as a command from
\texttt{/encs/pkg/netbeans-6.5.1/root/bin/netbeans}
on lab desktops, is used
for legacy SOA and BPEL exercises. The rest of the section describes
this version of NetBeans as needed to be set up for the exercises.
You can create an alias for yourself and place it e.g. in your \file{.cshrc},
to shorten it:

\begin{verbatim}
addams.mokhov [~] % alias netbeans651 /encs/pkg/netbeans-6.5.1/root/bin/netbeans
addams.mokhov [~] % netbeans651
\end{verbatim}



\subsection{Step-by-Step Environment Setup}
\label{sect:linux-env-setup}

\begin{enumerate}

\item
Login to Linux. If you never did before likely
your default qindow manager is GNOME.

\item
Open up the terminal: ``Applications'' $\rightarrow$ ``Accessories'' $\rightarrow$ ``Terminal''.
The window similar to \xf{fig:screenshot-terminal} should pop-up.

\begin{figure}[htpb]
	\centering
	\includegraphics[width=\textwidth]{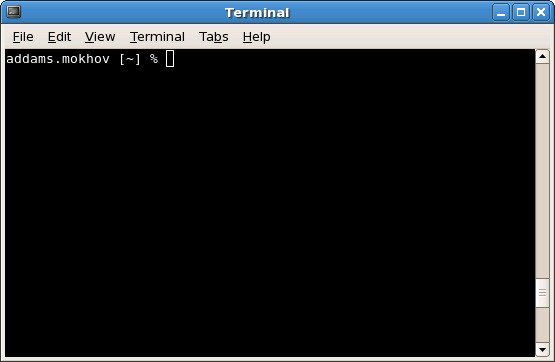}
	\caption{Terminal Window}
	\label{fig:screenshot-terminal}
\end{figure}

\item
Configure your Java 1.6 to be the default as outlined
in \xs{sect:java-16}, and an example is shown in \xf{fig:screenshot-terminal-java-16}.

\begin{figure}[htpb]
	\centering
	\includegraphics[width=\textwidth]{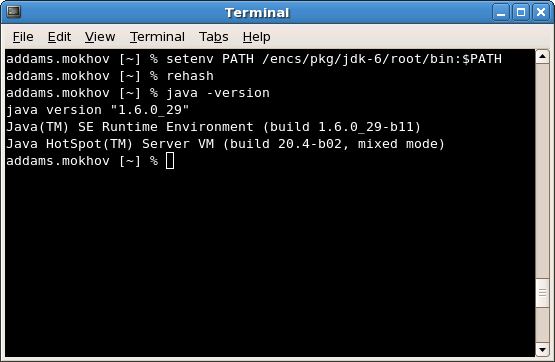}
	\caption{Setting up Java 1.6 as a Default in the Terminal}
	\label{fig:screenshot-terminal-java-16}
\end{figure}

\item
In {\em the same terminal window}, change your \api{HOME} environment variable
to that of your 1GB group directory. This will allow most portions of
NetBeans to write the temporary and configuration files there by default
instead of your main Unix home directory. I use an equivalent directory of mine
{\mygrpfull}\index{Files!{\mygrpfull}},
{\em as an example} -- and you should be using the directory assigned to you
with your group 1GB quota. An example to do so is very similar as to set up
\api{PATH}, except it is a single entry. It is exemplified in \xf{fig:screenshot-terminal-home}.
Unlike \api{PATH}, \api{HOME} is {\em not} recommended to be hardcoded to change
your \api{HOME} in \file{.cshrc}.

\begin{figure}[htpb]
	\centering
	\includegraphics[width=\textwidth]{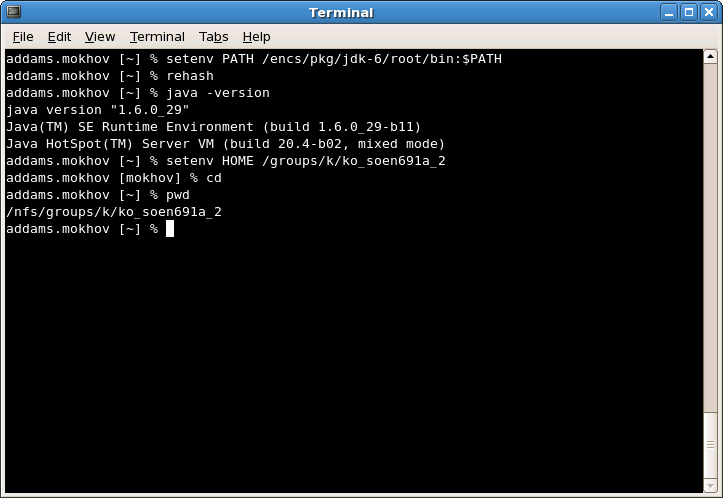}
	\caption{Setting up \api{HOME} to the Group Directory}
	\label{fig:screenshot-terminal-home}
\end{figure}

\item
Create the following directories in your new group \api{HOME} (your 1GB group directory):

\begin{verbatim}
mkdir .netbeans .netbeans-derby .netbeans-registration
chmod g+rwsx .netbeans .netbeans-derby .netbeans-registration
ls -al
\end{verbatim}

These directories will hold all the configuration and
deployment files pertaining to NetBeans, the Derby database,
and the domains for GlassFish'es operation.
The overall content may easily reach 80MB in total disk
usage for all these directories just to start up.

\item
Disk usage, quota, and big files (in case running out of space)
can be checked for using the following commands:
\begin{verbatim}
quota
du -h
bigfiles
\end{verbatim}

\item
In your {\em real home directory} (open another Terminal),
remove any previous NetBeans et co. setup
files you may have generated from the previous runs:

\file{.asadminpass}

\file{.asadmintruststore}

\file{.netbeans*}

\file{.personalDomain*}

\noindent
(assuming no important data for you are saved there):

\begin{verbatim}
\rm -rf .netbeans* .personalDomain* .asadmin*
\end{verbatim}

\item
In your {\em real home directory} create symbolic links (``shortcuts'')
to the same NetBeans directories now found in your group directiry
you made earlier. This is just in case you launch NetBeans without
redirecting the \api{HOME}, it still goes to the group directory
without impending your main quota:

\scriptsize
\begin{verbatim}
addams.mokhov [~] % pwd
/nfs/home/m/mokhov
addams.mokhov [~] % ln -s /groups/k/ko_soen691a_2/.netbeans* .
addams.mokhov [~] % ls -ld .netbeans*
lrwxrwxrwx 1 mokhov mokhov 33 Oct 31 18:37 .netbeans -> /groups/k/ko_soen691a_2/.netbeans
lrwxrwxrwx 1 mokhov mokhov 39 Oct 31 18:37 .netbeans-derby
    -> /groups/k/ko_soen691a_2/.netbeans-derby
lrwxrwxrwx 1 mokhov mokhov 46 Oct 31 18:37 .netbeans-registration
    -> /groups/k/ko_soen691a_2/.netbeans-registration
addams.mokhov [~] % 
\end{verbatim}
\normalsize

\item
In the {\em group home} terminal window launch NetBeans, by executing the
command \tool{netbeans~\&}, and after some time it should fully
start up {\em without} of any errors.

\item\label{item:netbeans691}
This is NetBeans 6.9.1, the 6.5.1 will look slightly different in some places.
It is covered at step \ref{item:netbeans651}.

You will be prompted to
allow Sun/Oracle to collect your usage information and register;
it is recommended to answer ``No'' to both. And then you will
see a left-hand-side (LHS) menu, the main editor page with the
default browsed info, and the top menu of the NetBeans,
as shown in \xf{fig:screenshot-netbeans-ide-691}.

\begin{figure}[htpb]
	\centering
	\includegraphics[width=\textwidth]{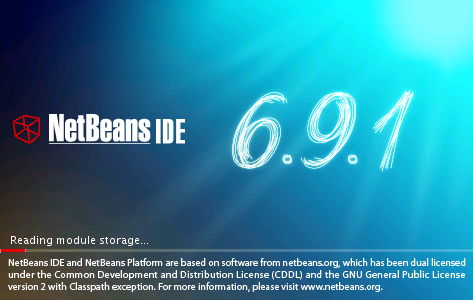}
	\caption{NetBeans 6.9.1 Starting Up}
	\label{fig:screenshot-starting-netbeans-ide}
\end{figure}

\begin{figure}[htpb]
	\centering
	\includegraphics[width=\textwidth]{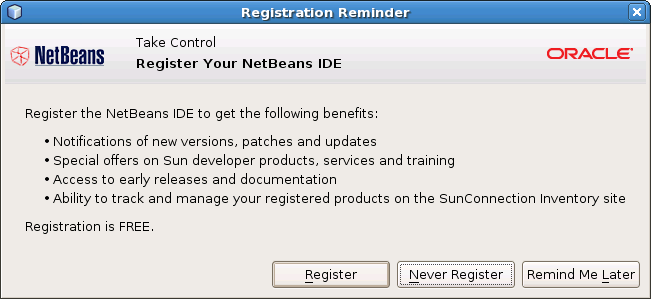}
	\caption{NetBeans 6.9.1 Registration}
	\label{fig:screenshot-registration-reminder}
\end{figure}

\begin{figure}[htpb]
	\centering
	\includegraphics[width=\textwidth]{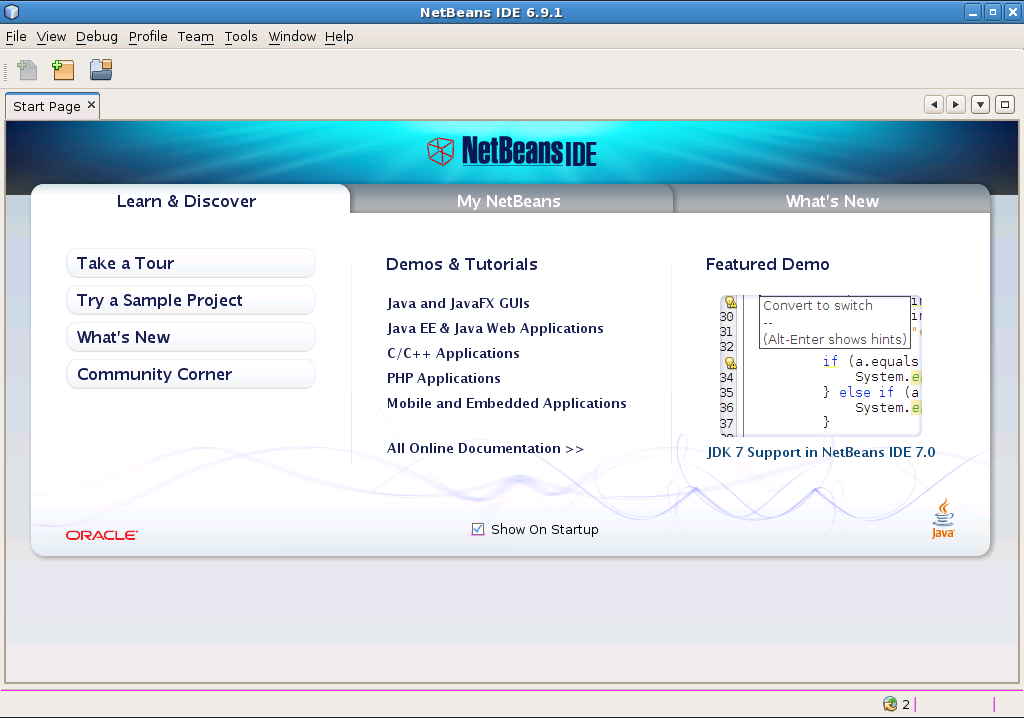}
	\caption{NetBeans 6.9.1 Start-up Screen}
	\label{fig:screenshot-netbeans-ide-691}
\end{figure}

\begin{enumerate}

\item
A: screenshot-warning-gf301-domain-mpass-691.png (8 OK)
\begin{figure}[htpb]
	\centering
	\includegraphics[width=\textwidth]{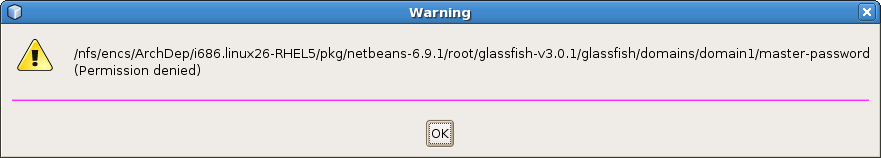}
	\caption{screenshot-warning-gf301-domain-mpass-691}
	\label{fig:screenshot-warning-gf301-domain-mpass-691}
\end{figure}

\item
B: screenshot-warning-gf3-prelude-651.png
\begin{figure}[htpb]
	\centering
	\includegraphics[width=\textwidth]{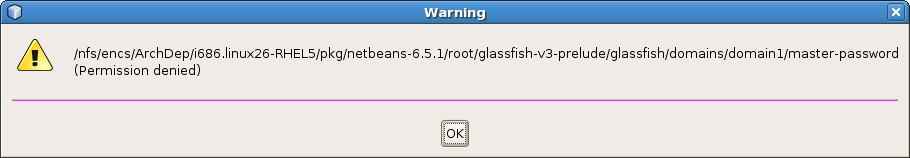}
	\caption{screenshot-warning-gf3-prelude-651}
	\label{fig:screenshot-warning-gf3-prelude-651}
\end{figure}

\item
\begin{figure}[htpb]
	\centering
	\includegraphics[width=\textwidth]{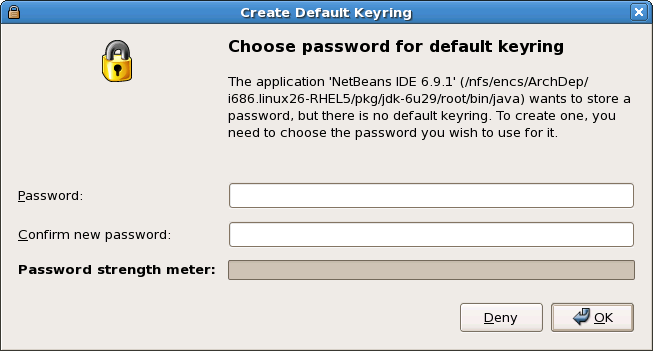}
	\caption{screenshot-create-default-keyring}
	\label{fig:screenshot-create-default-keyring}
\end{figure}

\item
\begin{figure}[htpb]
	\centering
	\includegraphics[width=\textwidth]{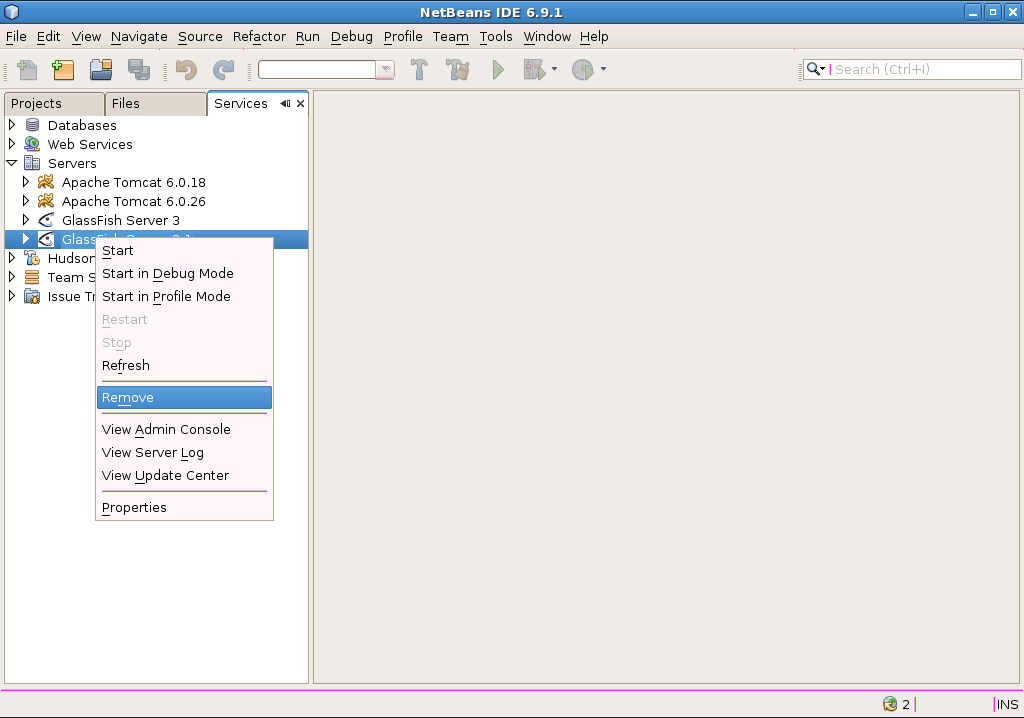}
	\caption{screenshot-netbeans691-remove-gf31}
	\label{fig:screenshot-netbeans691-remove-gf31}
\end{figure}

\item
\begin{figure}[htpb]
	\centering
	\includegraphics[width=\textwidth]{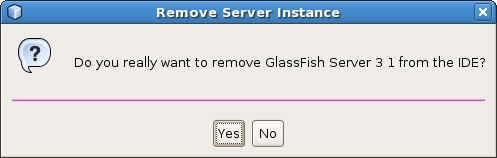}
	\caption{screenshot-remove-server-instance-gf31}
	\label{fig:screenshot-remove-server-instance-gf31}
\end{figure}
 

\item
\begin{figure}[htpb]
	\centering
	\includegraphics[width=\textwidth]{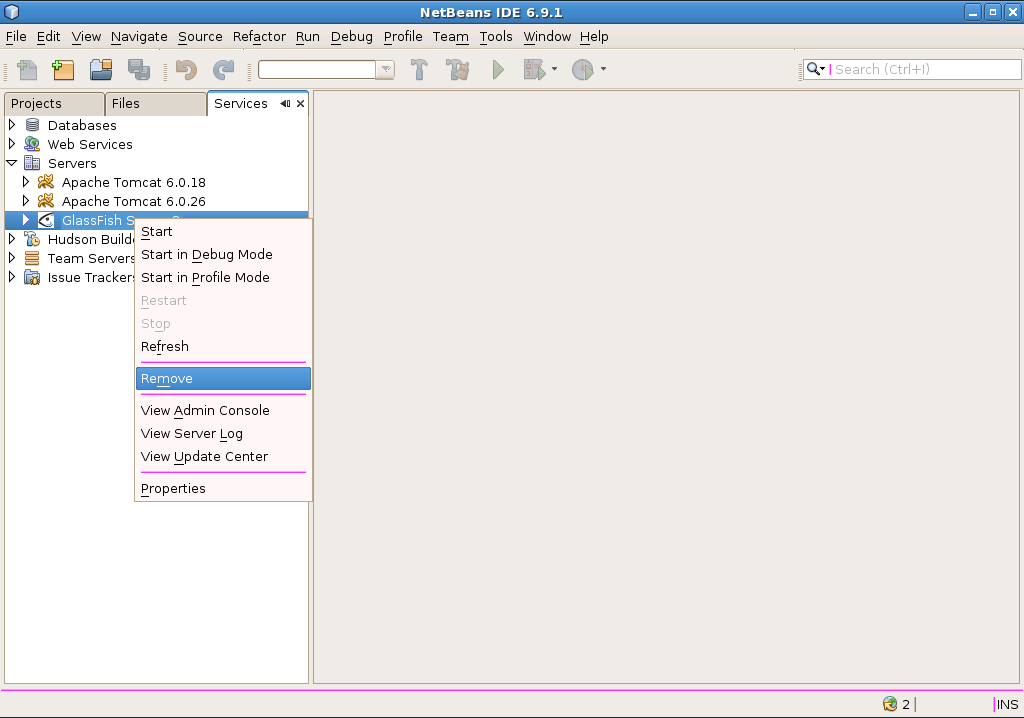}
	\caption{screenshot-netbeans691-remove-gf3}
	\label{fig:screenshot-netbeans691-remove-gf3}
\end{figure}

\item
\begin{figure}[htpb]
	\centering
	\includegraphics[width=\textwidth]{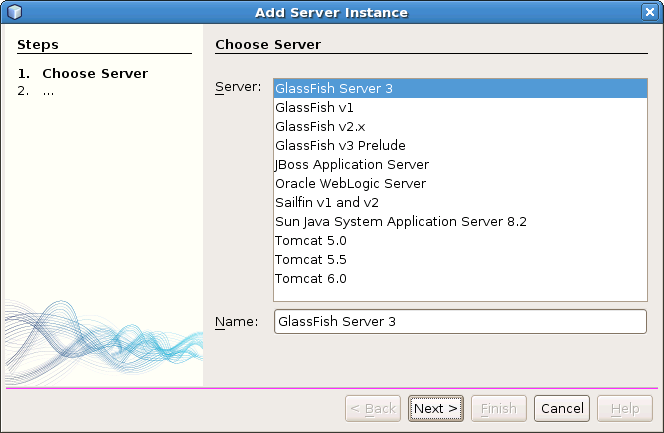}
	\caption{screenshot-add-server-instance-gf3}
	\label{fig:screenshot-add-server-instance-gf3}
\end{figure}

\item
\begin{figure}[htpb]
	\centering
	\includegraphics[width=\textwidth]{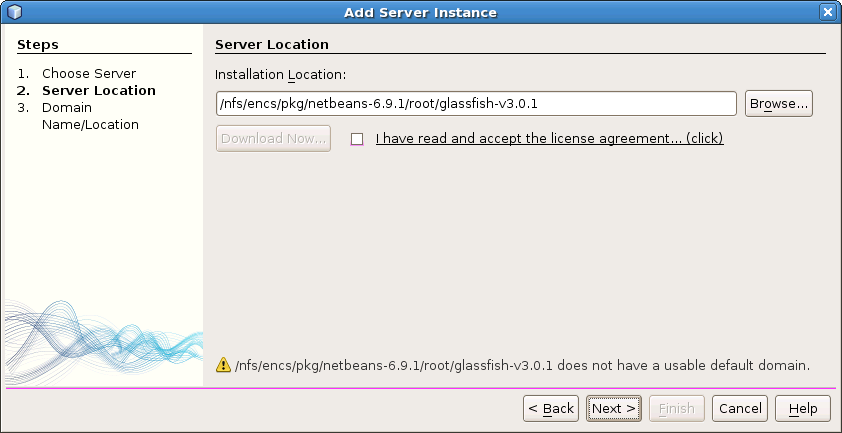}
	\caption{screenshot-add-server-instance-gf301-path}
	\label{fig:screenshot-add-server-instance-gf301-path}
\end{figure}

\item
\begin{figure}[htpb]
	\centering
	\includegraphics[width=\textwidth]{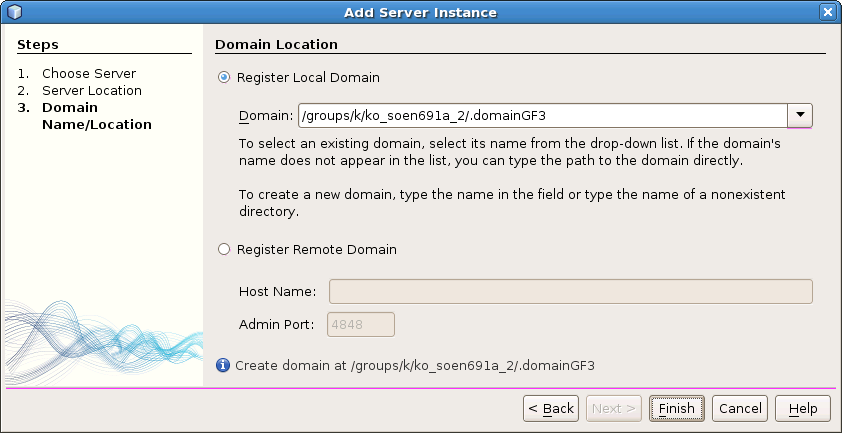}
	\caption{screenshot-add-server-instance-gf3-domain}
	\label{fig:screenshot-add-server-instance-gf3-domain}
\end{figure}

\item
\begin{figure}[htpb]
	\centering
	\includegraphics[width=\textwidth]{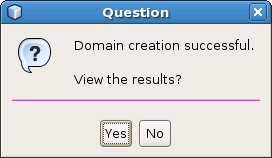}
	\caption{screenshot-domain-creation-results}
	\label{fig:screenshot-domain-creation-results}
\end{figure}

\item
\begin{figure}[htpb]
	\centering
	\includegraphics[width=\textwidth]{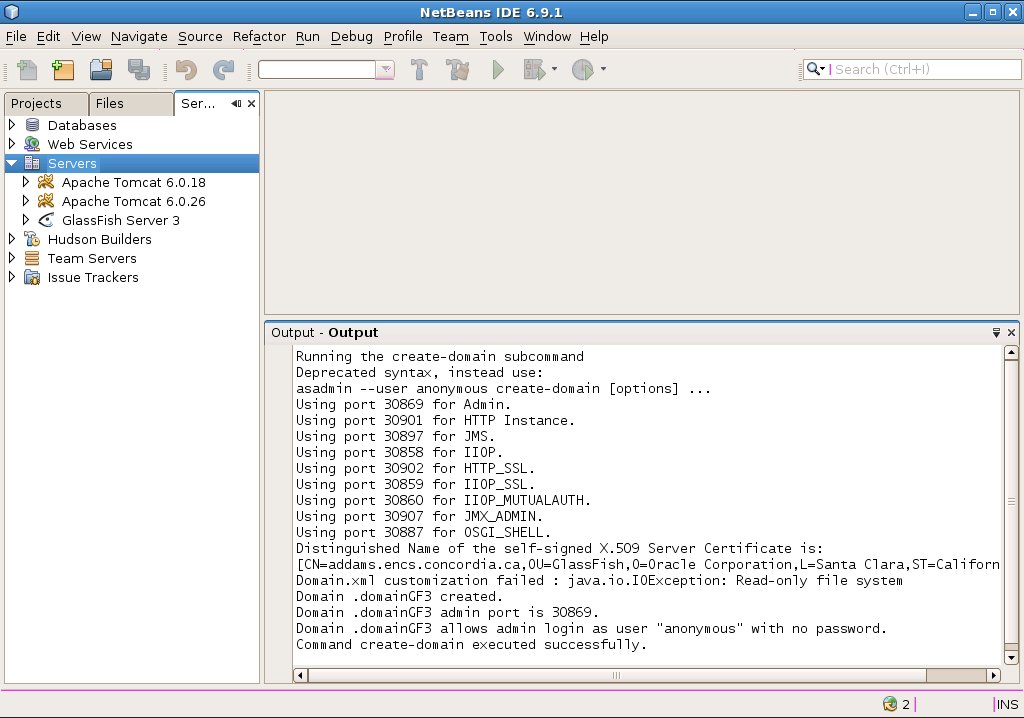}
	\caption{screenshot-netneans691-gf3-created}
	\label{fig:screenshot-netneans691-gf3-created}
\end{figure}

\end{enumerate}

\clearpage

\item\label{item:netbeans651}
This is NetBeans 6.5.1, the 6.9.1 will look slightly different in some places.
It is covered at step \ref{item:netbeans691}.

You will be prompted to
allow Sun/Oracle to collect your usage information and register;
it is recommended to answer ``No'' to both. And then you will
see a left-hand-side (LHS) menu, the main editor page with the
default browsed info, and the top menu of the NetBeans,
as shown in \xf{fig:screenshot-netbeans-ide-651}.

\begin{figure}[htpb]
	\centering
	\includegraphics[width=\textwidth]{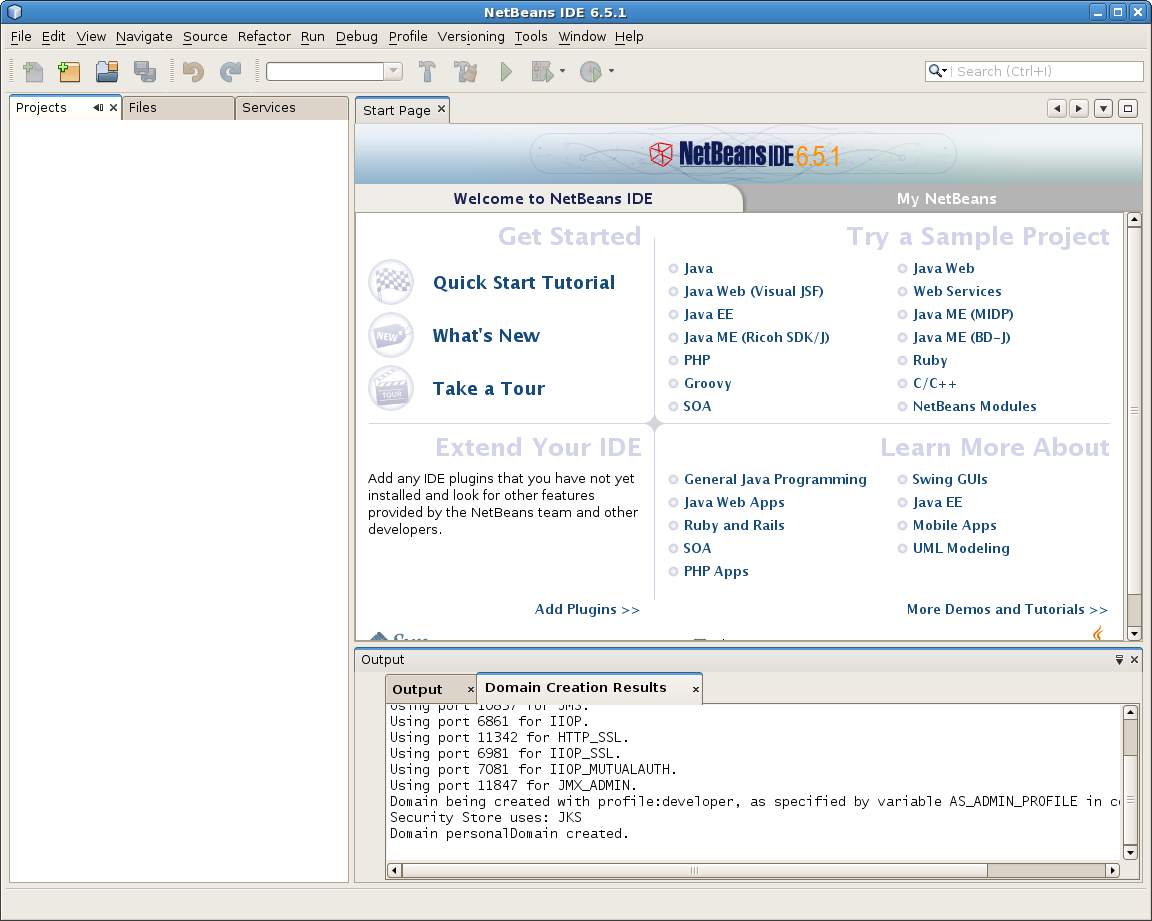}
	\caption{NetBeans 6.5.1 Start-up Screen}
	\label{fig:screenshot-netbeans-ide-651}
\end{figure}

\begin{enumerate}
\item
Navigate to the ``Services'' tab and expand the ``Server'' tree in the LHS
menu. You should be able to see a ``GlassFish V2'' entry there (among other things),
as shown in \xf{fig:screenshot-netbeans-services-servers}.

\begin{figure}[htpb]
	\centering
	\includegraphics[width=\textwidth]{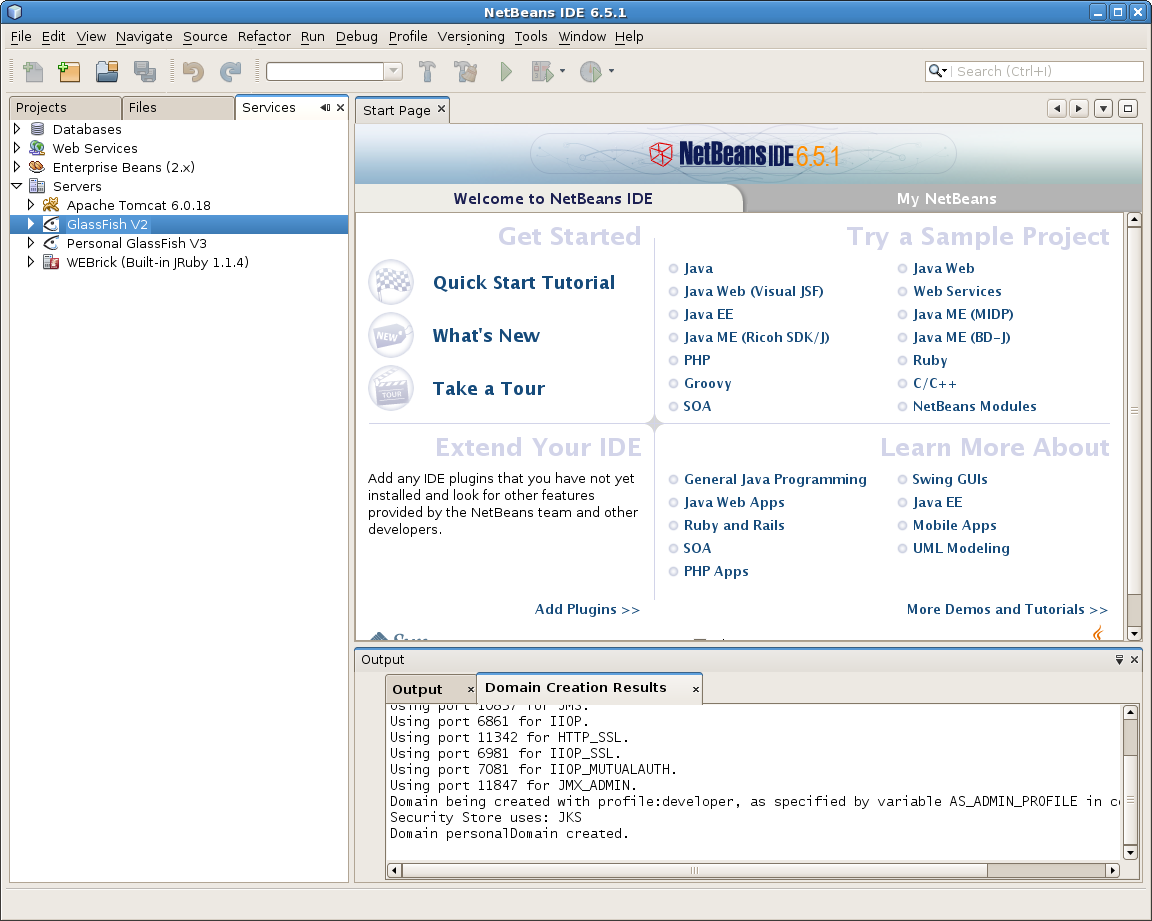}
	\caption{NetBeans: Services $\rightarrow$ Server $\rightarrow$ GlassFish V2}
	\label{fig:screenshot-netbeans-services-servers}
\end{figure}

\item
Right-click on ``GlassFish V2'' and then ``Properties'', as in \xf{fig:screenshot-servers-glassfish-properties}.
Observe the ``Domains folder'' and ``Domain Name''. If the folder points within
your normal home directory, you have to change it as follows (and then remove it from your
personal home directory):

\begin{figure}[htpb]
	\centering
	\includegraphics[width=\textwidth]{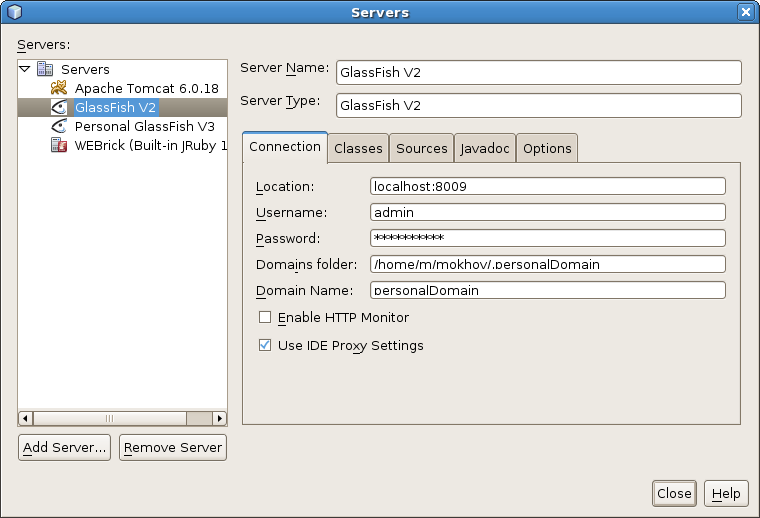}
	\caption{Right-click GlassFish V2 $\rightarrow$ Properties}
	\label{fig:screenshot-servers-glassfish-properties}
\end{figure}

	\begin{enumerate}
	\item
	Close the properties window.
	\item
	Right-click on ``GlassFish V2'' and then ``Remove''. Confirm the removal with ``Yes''.
	(You may as well remove ``Personal GlassFish V3'').
	\item
	Right-click on ``Servers'' and then ``Add Server...''.
	\item
	Select ``GlassFish V2'' and then ``Next'', and ``Next''.
	\item
	Then, for the ``Domain Folder Location'' ``Browse'' or paste
	your group directory followed by an appended domain name,
	e.g. \file{/groups/r/rm\_soen691a\_4/domainGF2} in my case,
	notice where \file{domainGF2} is an arbitrary name of a directory under your
	group directory that is not existing yet, give it any name you like, and then
	press ``Next''.
	\item
	\label{step:credentials}
	Pick a user name and a password for the admin console (web-based) of GlassFish.
	The NetBeans default (of the GlassFish we removed) is `admin' and `adminadmin'.
	It is {\em strongly} suggested however you do {\em NOT} follow the default,
	and pick something else. Do {\em NOT} make it equal to your ENCS account either.
	\item
	``Next'' and ``Finish''. Keep the ports at their defaults {\bf EXCEPT}
	set HTTP port to 8085 and HTTPS to 8185. Notice it
	may take time to restart the new GlassFish instance and recreate
	your domain you indicated in the group folder.
	\end{enumerate}

\item
Right-click on ``GlassFish V2'' again and select ``Start''. It may also take
some time to actually start GlassFish; watch the bottom-right corner
as well as the output window for the startup messages and status.
There should be no errors. Apache Derby service should
have started.

\item
Once started, right-click on ``GlassFish V2'' again, and select ``View Admin Console''.
You should see the GlassFish login window pop-up in the Firefox web browser,
looking as shown in \xf{fig:screenshot-firefox-glassfish-login}.

\begin{figure}[htpb]
	\centering
	\includegraphics[width=\textwidth]{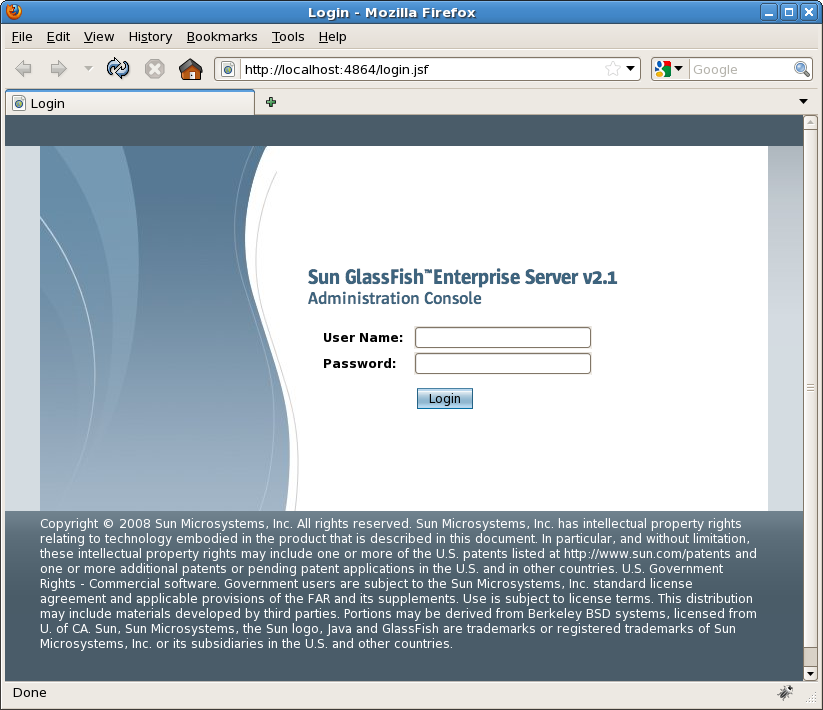}
	\caption{GlassFish Admin Console Login Screen}
	\label{fig:screenshot-firefox-glassfish-login}
\end{figure}

\item
To log in, use the username and password you created earlier in Step~\ref{step:credentials}.

\item
In your {\em group home} terminal, download additional libraries from {\coursewww}
(formperly from \cite{bpelse}). In the lab, you will
only need 2 (\file{wsdlextlib.jar} and \file{saxonlib.jar}) out of typical 3,
because the version installed in ENCS already includes
the 3rd (\file{bpelserviceengine.jar}). You will likely need the 3rd file however,
for your laptop or home desktop in Windows.
%

\item
In your GlassFish console web page, under
``Common Tasks'' $\rightarrow$ ``JBI'' $\rightarrow$ ``Shared Libraries''
you need to install the two libraries we downloaded (3 for your Windows
laptop or home desktop) by clicking ``Install'' and following the steps
by browsing to the directory where you downloaded the files and installing them.
Then, once installed \texttt{sun-saxon-library} and \texttt{sun-wsdl-ext-library}
should be listed under the ``Shared Libraries''.

You can also perform this step within NetBeans itself, by expanding the
``GlassFish V2'' tick, and then ``JBI'', then right-clicking on 
``Shared Libraries'' $\rightarrow$ ``Install.

It is {\em imperative} if you are using NetBeans 6.7.1 on your own
systems, to install these libraries first before moving onto the
next step and installing the BPEL engine.

\item
Make sure under ``Components'' you have \texttt{sun-bpel-engine}.
Linux boxes in the labs should have it installed with the NetBeans,
at home it's the 3rd file -- \file{bpelserviceengine.jar}, that
may need to be installed using the similar procedure as in the
previous step. Roughly, how your ``Components'' and ``Shared Libraries''
should look like is in \xf{fig:screenshot-glassfish-firefox-libraries-components}.

Similarly to the previous step, the installation of this jar
can be done within NetBeans under the ``Service Engines'' subtree
instead of ``Shared Libraries''.

\begin{figure}[htpb]
	\centering
	\includegraphics[angle=90,width=\textwidth]{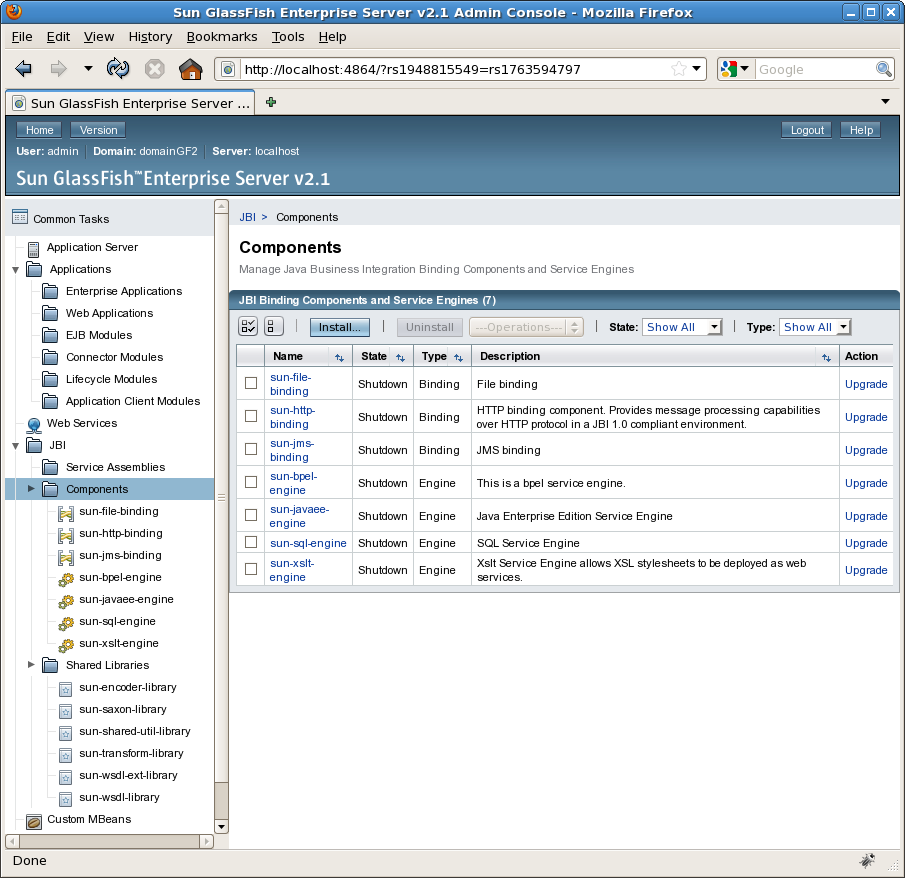}
	\caption{List of Components and Shared Libraries Installed in GlassFish}
	\label{fig:screenshot-glassfish-firefox-libraries-components}
\end{figure}
\end{enumerate}

\end{enumerate}

On this the NetBeans 
environment setup should be complete.
You will technically not need to repeat except if you
remove all the files from your group directory.

\section{Step-by-Step Simple Application and Web Service Creation and Testing}
\label{sect:simple-ws-app-testing}

\subsection{NetBeans 6.5.1}

\begin{enumerate}
\item
Go to the ``Projects'' tab in NetBeans.
\item
Then ``File'' $\rightarrow$ ``New Project''.
\item
Choose ``Java EE'' $\rightarrow$ ``Enterprise Application'',
as shown in \xf{fig:screenshot-new-project-j2ee-ea},
and then ``Next''.

\begin{figure}[htpb]
	\centering
	\includegraphics[width=\textwidth]{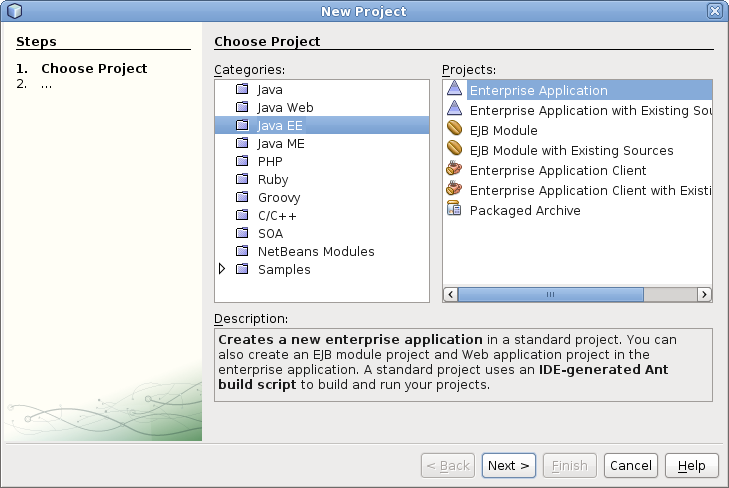}
	\caption{``Java EE'' $\rightarrow$ ``Enterprise Application''}
	\label{fig:screenshot-new-project-j2ee-ea}
\end{figure}

\item
Give the project properties, like Project Name to be ``A1'',
project location somewhere in your group directory, e.g.
as for me shown in \xf{fig:screenshot-new-ea-location},
and then ``Next''.

\begin{figure}[htpb]
	\centering
	\includegraphics[width=\textwidth]{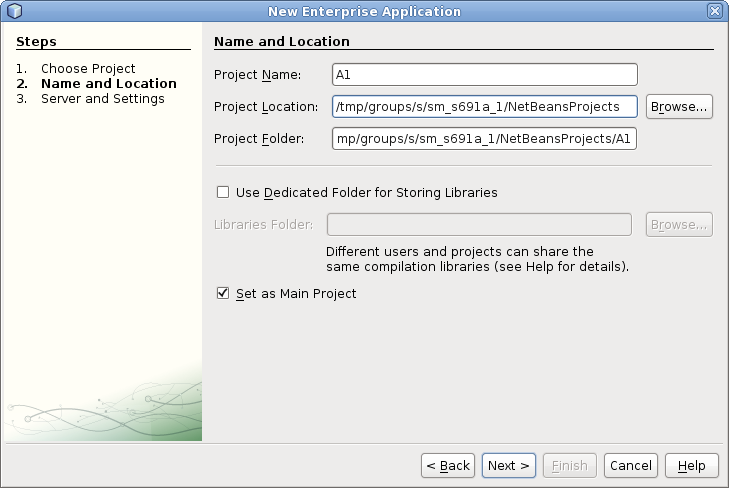}
	\caption{NetBeans Programming Projects Location}
	\label{fig:screenshot-new-ea-location}
\end{figure}

\item
In the next tab, you can optionally enable ``Application Client Module''
for an example, and keep the rest at their defaults, e.g. as shown in
\xf{fig:screenshot-new-ea-server-client}. Notice, I altered the client
package \api{Main} class to be in \api{soen691a.a1.Main}. It is not
strictly required in here as you can test your web services using
web service unit testing tools built-into the IDE.

\begin{figure}[htpb]
	\centering
	\includegraphics[width=\textwidth]{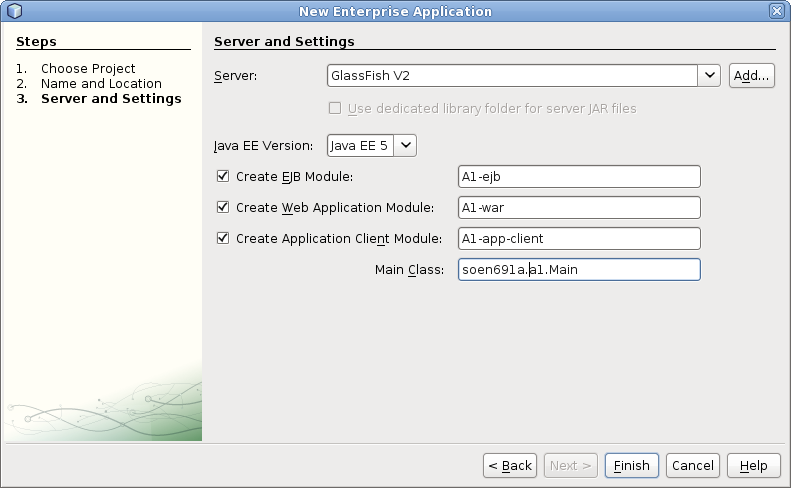}
	\caption{A1's Example Server and Client Settings}
	\label{fig:screenshot-new-ea-server-client}
\end{figure}

\item
Click ``Finish'' to create your first project with the above settings.
You should see something that looks like as shown in \xf{fig:screenshot-a1-project-created},
after some of the tree elements expanded.

\begin{figure}[htpb]
	\centering
	\includegraphics[angle=90,width=\textwidth]{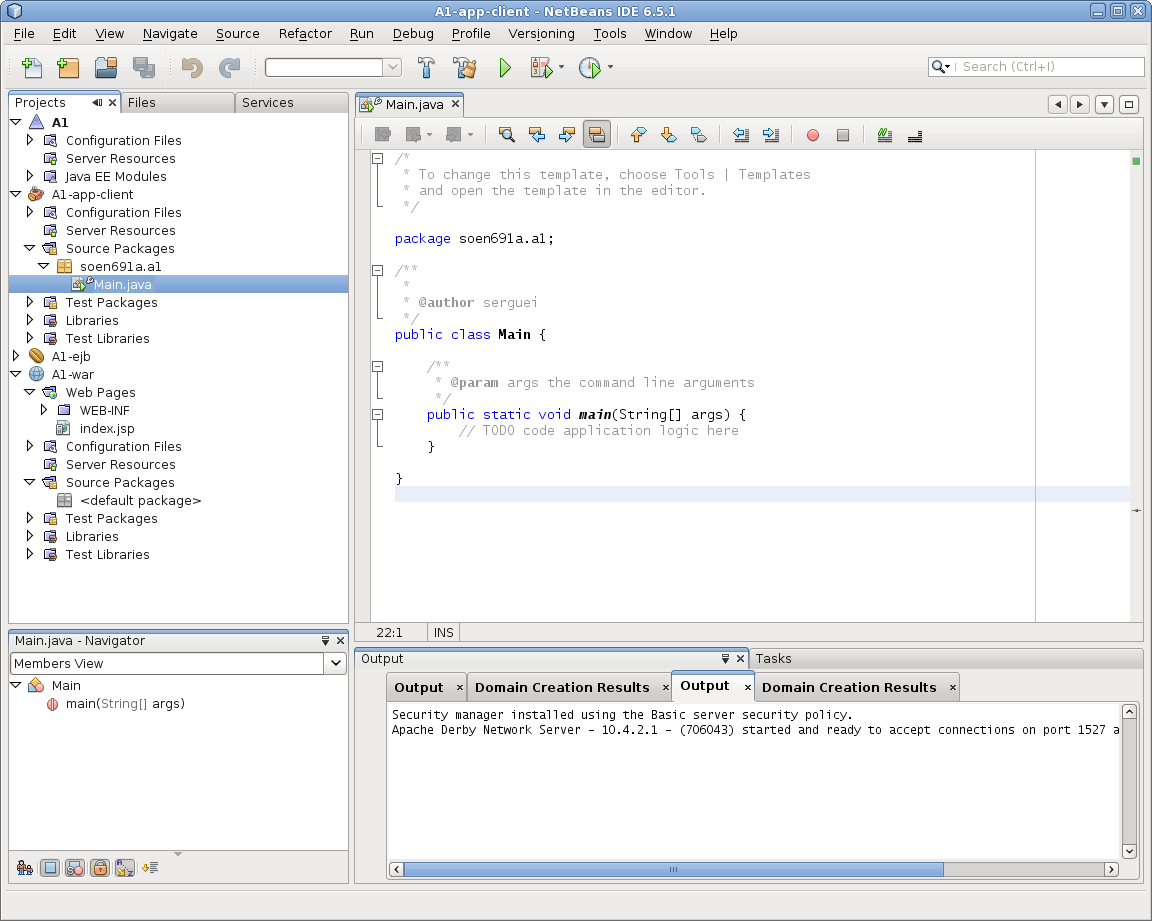}
	\caption{A1 Project Tree}
	\label{fig:screenshot-a1-project-created}
\end{figure}

\item
Under \api{A1-war}, create a package, called \api{soen691a}
by right-clicking under ``A1'' $\rightarrow$ ``Source Packages'' $\rightarrow$ ``New'' $\rightarrow$ ``Java Package'' $\rightarrow$ ``Package Name'': \api{soen691a}. Then ``Finish''.

\item
Create a ``Web Service'' under that package, by
right-click on the newly created
package $\rightarrow$ ``New'' $\rightarrow$ ``Web Service'' $\rightarrow$ ``Web Service Name'' $\rightarrow$  \api{Login},
as shown in \xf{fig:screenshot-new-ws-login}.

\begin{figure}[htpb]
	\centering
	\includegraphics[width=\textwidth]{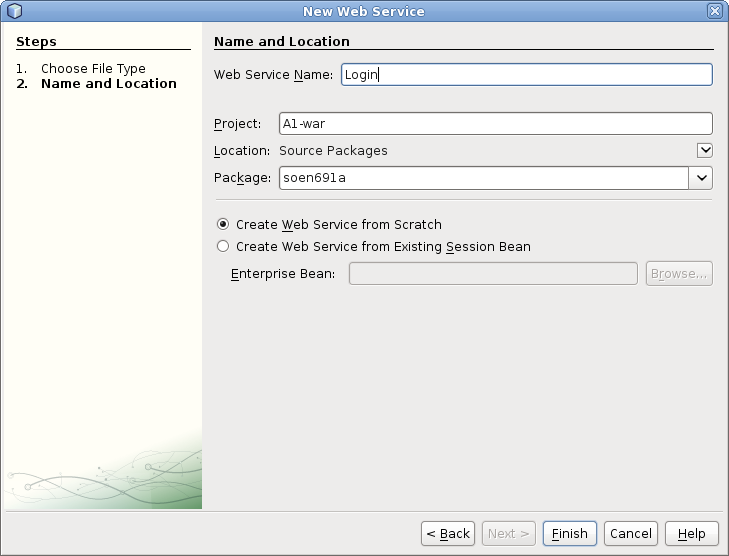}
	\caption{New \api{Login} Web Service}
	\label{fig:screenshot-new-ws-login}
\end{figure}

\item
The LHS project tree if expanded would look like shown in \xf{fig:screenshot-new-ws-login}.
\begin{figure}[htpb]
	\centering
	\includegraphics[angle=90,width=\textwidth]{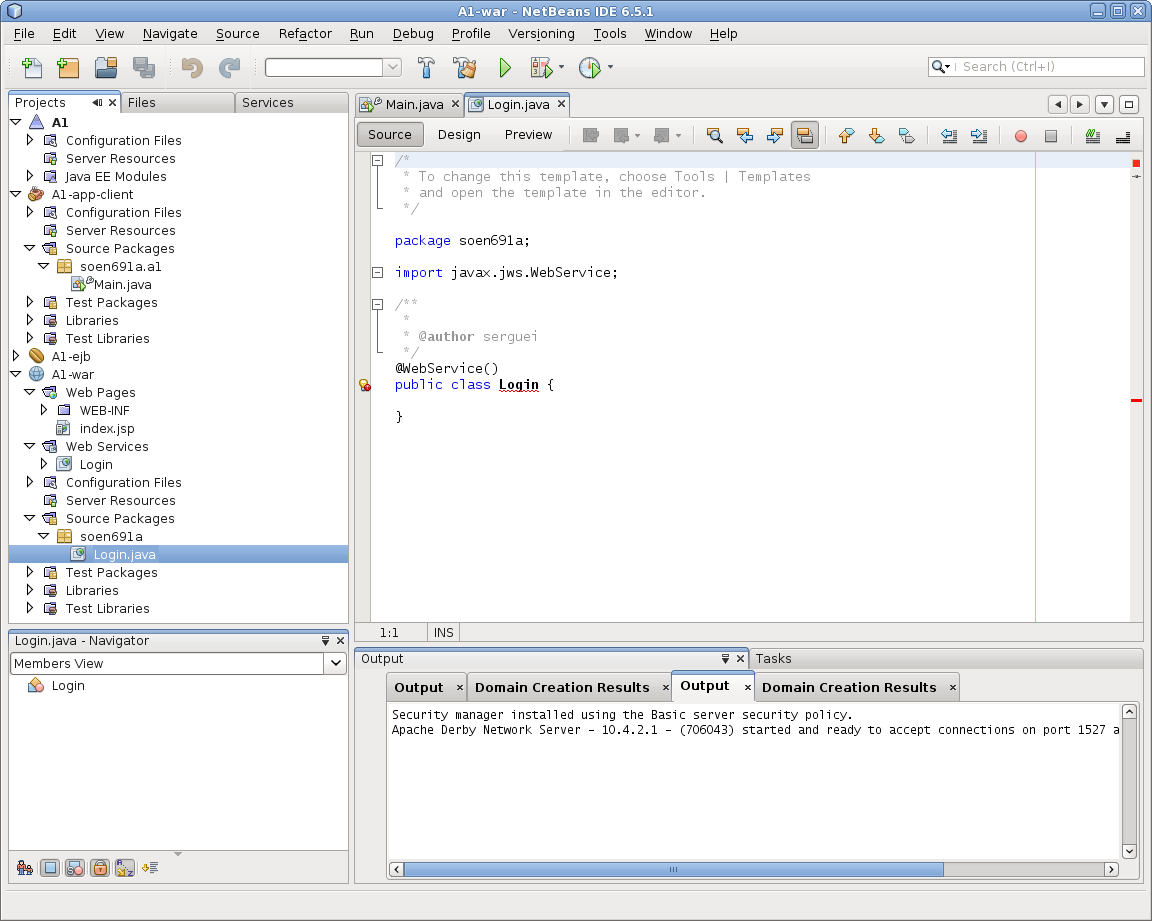}
	\caption{A1 Project Tree after Login Web Service Creation}
	\label{fig:screenshot-a1-login-ws-files}
\end{figure}

\item
Right-click on \api{Login} WS, and select ``Add Operation...'' and
create a web method \api{login()}, as shown in \xf{fig:screenshot-add-operation-login}.

\begin{figure}[htpb]
	\centering
	\includegraphics[width=.7\textwidth]{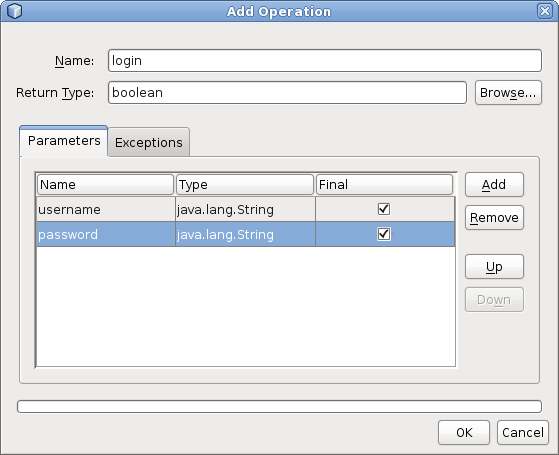}
	\caption{Adding a Web Method \api{login()}}
	\label{fig:screenshot-add-operation-login}
\end{figure}

\item
After the web method \api{login()} appears as a stub inside the \api{Login}
class with \texttt{return false;} by default. For quick unit testing
of the new method, implement it with some test user name and password
as shown in \xf{fig:screenshot-netbeans-ide-login-web-method},
which will later be replaced to be read from the XML file.

\begin{figure}[htpb]
	\centering
	\includegraphics[angle=90,width=\textwidth]{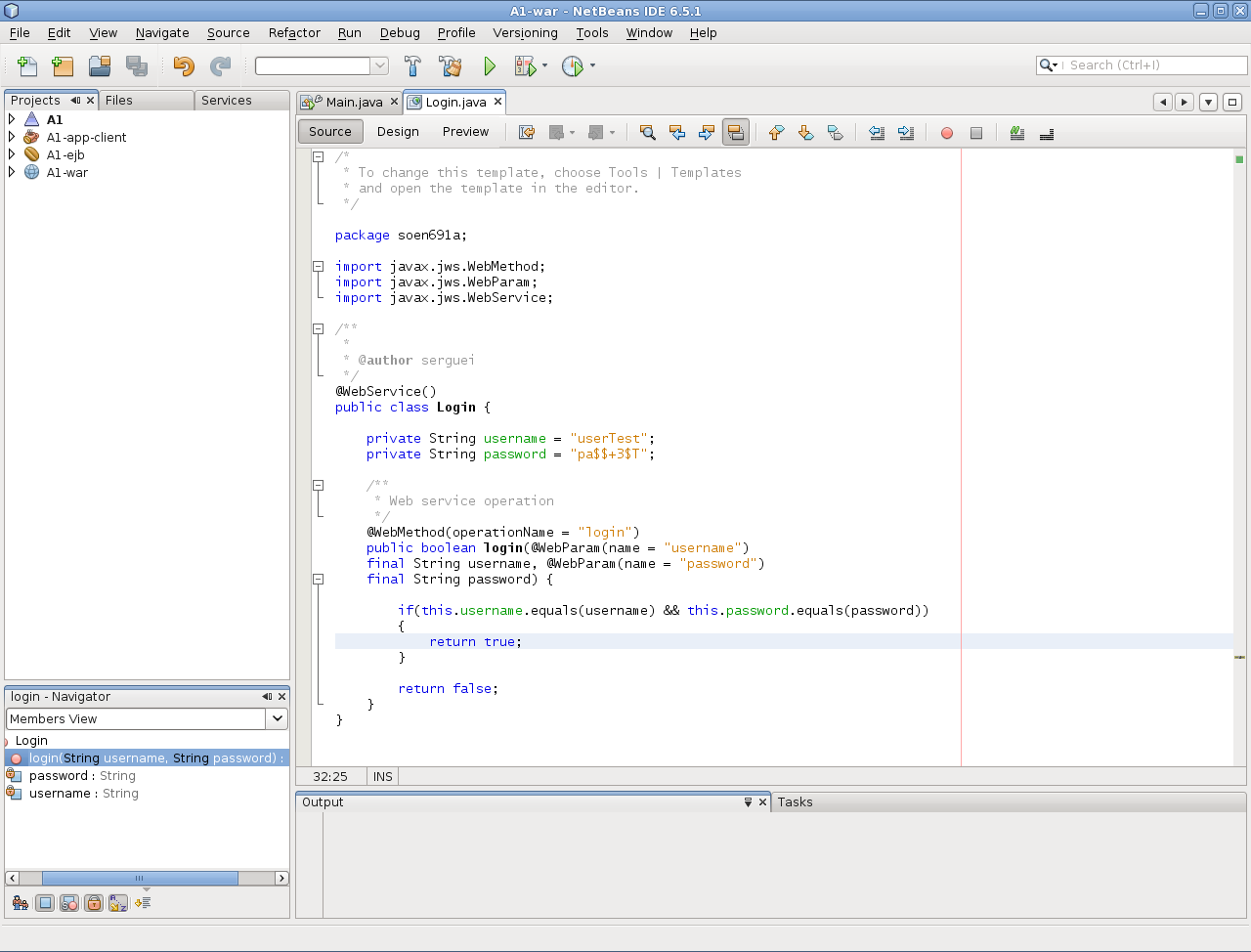}
	\caption{Implementing a Simple Web \api{login()} Method for Quick Unit Testing}
	\label{fig:screenshot-netbeans-ide-login-web-method}
\end{figure}

\item
Perform a simple unit test for the web method. Your GlassFish must be running
and you have to ``start'' your project by deploying  -- just press the green
angle ``play'' button. You should see a ``Hello World'' page appearing in
your browser.

\item
Then, under ``A1-war'' $\rightarrow$ ``Web Services'' $\rightarrow$ ``Login'' right-click
on \api{Login} and select ``Test Web Service''. It should pop-up
another browser window (or tab) titled something like ``LoginService Web Service Tester''
with a pre-made form to test inputs to your web method(s), as shown in
\xf{fig:screenshot-login-service-ws-test-firefox}.

\begin{figure}[htpb]
	\centering
	\includegraphics[width=\textwidth]{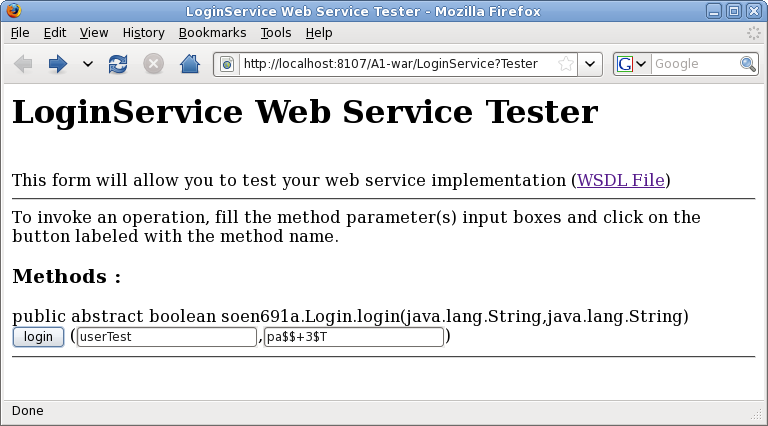}
	\caption{Unit-testing Page for the \api{Login} WS}
	\label{fig:screenshot-login-service-ws-test-firefox}
\end{figure}

\item
Fill-in the correct test values
that we defined earlier for login
and press the ``login'' button.
Observe the exchanged SOAP XML messages
and the \api{true} value returned as a result,
as shown in \xf{fig:screenshot-login-method-invocation-firefox}.

Then try any wrong combination
of the username and password and see that it
returns \api{false}. This completes basic
verification of your web service -- that is can
be successfully deployed and ran, and its method(s)
unit-tested on the page.

\begin{figure}[htpb]
	\centering
	\includegraphics[totalheight=\textheight]{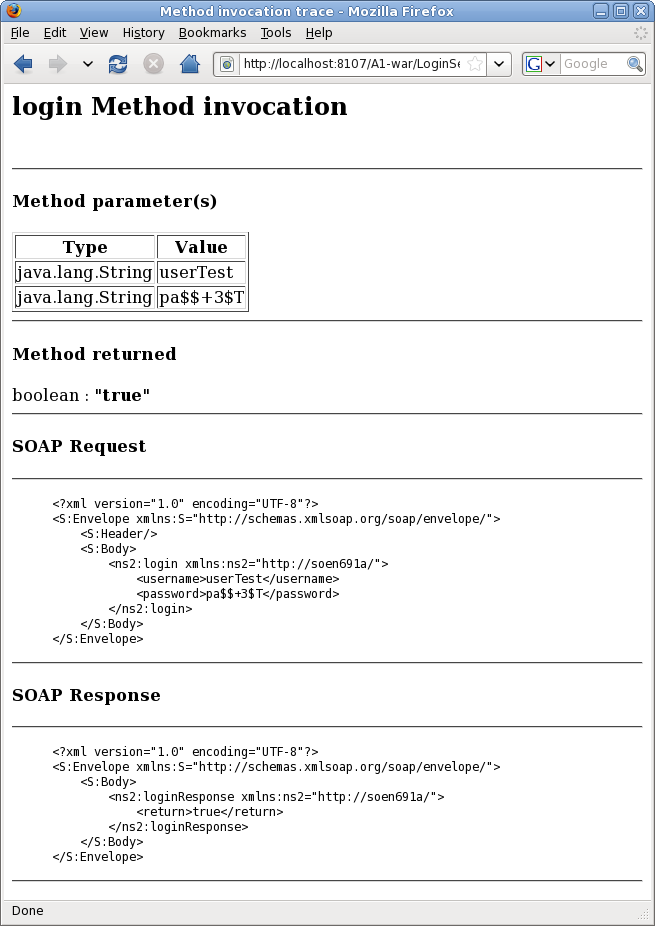}
	\caption{\api{login()} Web Method Invocation Trace}
	\label{fig:screenshot-login-method-invocation-firefox}
\end{figure}

\item
Java-based client callee of a web service has to
be defined e.g. as a WS client, as shown in
earlier screenshots as ``A1-app-client'', which
has a \api{Main.main()} method. In that method
you simply invoke the desired service by calling
its web method after a number of instantiations.
It may look like you are calling a local method
of a local class, but, in fact, on the background
there is a SOAP message exchange, marshaling/demarshaling
of data types, etc. and actually connection to a web
service, posting a request, receiving and parsing
HTTP response, etc. all done by the middleware.

Steps:

\begin{enumerate}
\item
Right-click ``A1-app-client'' $\rightarrow$ ``New'' $\rightarrow$ ``Web Service Client''.
A dialog shown in \xf{fig:screenshot-new-ws-client} should appear. Click ``Browse''.

\begin{figure}[htpb]
	\centering
	\includegraphics[width=\textwidth]{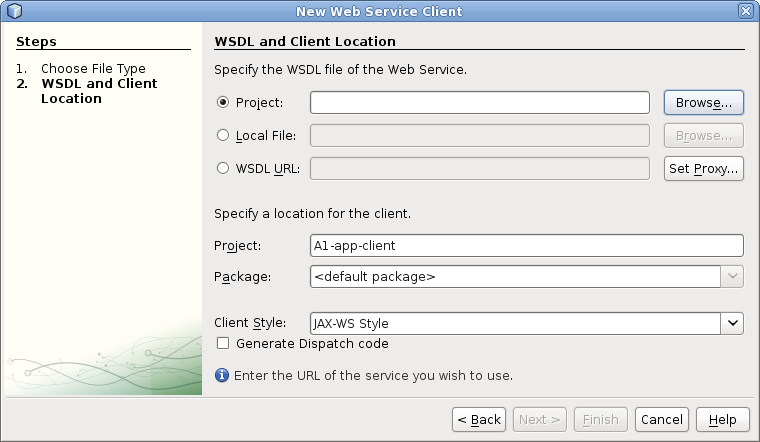}
	\caption{Creating a New Web Services Client in the Client Application Package from a Project}
	\label{fig:screenshot-new-ws-client}
\end{figure}

\item
Select your web service to generate a reference client for, as
e.g. shown in \xf{fig:screenshot-browse-ws} and click ``OK''.

\begin{figure}[htpb]
	\centering
	\includegraphics[width=.5\textwidth]{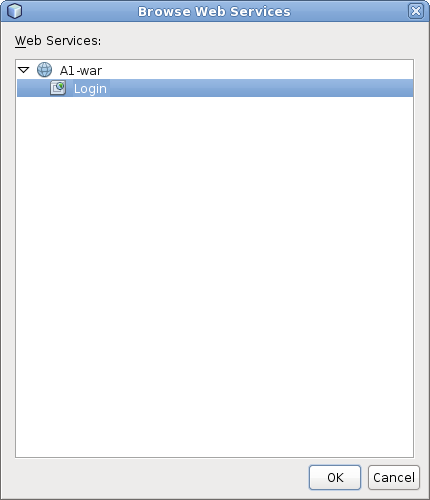}
	\caption{Selecting the Service to Create a Client For from the Project}
	\label{fig:screenshot-browse-ws}
\end{figure}

\item
Having selected the service to generate the WS client code for,
you should see the URL, as shown in \xf{fig:screenshot-new-ws-client-project}
``Finish'', re-deploy (green ``Play'' button).

\begin{figure}[htpb]
	\centering
	\includegraphics[width=\textwidth]{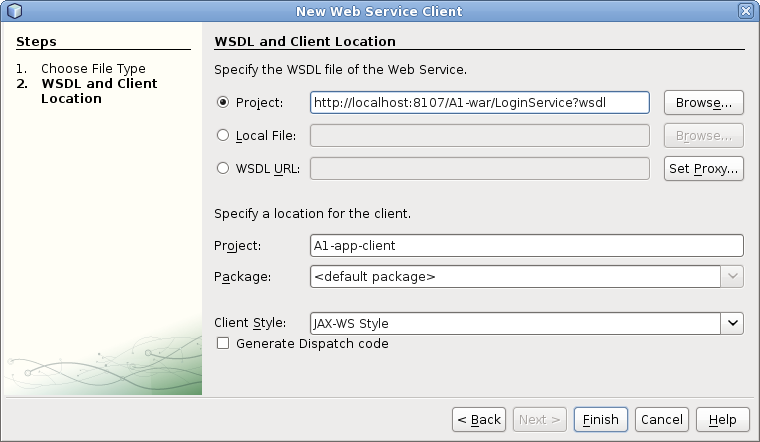}
	\caption{Creating a New Web Services Client Nearly Done. Notice the URL}
	\label{fig:screenshot-new-ws-client-project}
\end{figure}


\item
Then, in \api{Main}, import the generated code classes to invoke the
service, as shown in \xl{list:main-java-based-callee}.

\begin{lstlisting}[
    label={list:main-java-based-callee},
    caption={Invoking a Web Service from a Plain Java Class},
    style=codeStyle
    ]
package soen691a.a1;

import soen691a.Login;
import soen691a.LoginService;

/**
 * @author serguei
 */
public class Main {

    /**
     * @param args the command line arguments
     */
    public static void main(String[] args) {
        LoginService service = new LoginService();
        Login login  = service.getLoginPort();

        //...
        // Must be false
        boolean success = login.login("wrongusername", "wrongpasword");
        // Must be false
        success = login.login("wrongusername", "pa$$+3$T");
        // Must be false
        success = login.login("userTest", "wrongpasword");
        // Must be true
        success = login.login("userTest", "pa$$+3$T");
        //...
    }
}
\end{lstlisting}

\end{enumerate}

See also an example from {\dmarf}~\cite{dmarf06}.
%

\item
Sometimes your ports for HTTP and HTTPS can be different
from your home machine or multiple installations and a
lab machine. You can synchronize the ports in the client
side by expanding ``Web Service References'' and the
first node of a service in question, e.g. ``LoginService'',
then doing right-click it $\rightarrow$
``Edit Web Service Attributes'' $\rightarrow$
``Wsimport Options'' $\rightarrow$
``wsdlLocation'' -- and fix the port number in the URL.
Then click ``OK'', and ``Clean and Build'' (and ``Deploy'' if necessary)
the whole client project.

\item
Relative path for loading XML can be found
using \api{System.getProperty(``user.dir'')} to find out your
current working directory of the application, which is actually
relative to the \file{config/} subdirectory in your personal domain folder,
so it would be based on your deployment, but roughly:

\begin{verbatim}
System.getProperty("user.dir") + "../generated/....../users.xml"
\end{verbatim}

where ``\texttt{......}'' is the path leading to where your \file{users.xml}
and others actually are. You can configure Ant's \file{build.xml}
(actually \file{build-impl.xml} and other related files for
deployment to copy your XML data files into \file{config/}
automatically.

\item
Loading and querying XML with SAX 
is exemplified in \api{TestNN}
with {\marf}~\cite{marf-test-nn,marf}, specifically
at these CVS URLs:

\url{http://marf.cvs.sf.net/viewvc/marf/apps/TestNN/}

\url{http://marf.cvs.sf.net/viewvc/marf/marf/src/marf/Classification/NeuralNetwork/}

Do not validate your XML unless you specified a DTD
schema (not necessary here), just make sure your
tags are matching, properly nested, and closed.


\end{enumerate}

%
%
%
%
%
%
%

\section{Conclusion}

Please direct any problems and errors with these notes or any other
constructive feedback to \url{mokhov@cse.concordia.ca}.

\subsection{See Also}

\begin{itemize}
\item
GlassFish website~\cite{glassfish}.
\item
Unix/Linux commands~\cite{mokhov-unix-commands}.
\item
ENCS help: \url{http://www.encs.concordia.ca/helpdesk/}.
\item
An example of the XML parsing application, \api{TestNN}
with {\marf}~\cite{marf-test-nn,marf} using the built-in SAX parser.
\end{itemize}

\section{Acknowledgments}

\begin{itemize}
	\item Open-source community
	\item
		Faculty of Engineering and Computer Science,
		Department of Computer Science and Software Engineering,
		Concordia University, Montreal, Canada
	\item Ludeng Zhao
	\item Stephen Jiang
	\item SourceForge.net
\end{itemize}

\bibliographystyle{alpha}
\bibliography{service-oriented-arch-tutorial-notes}

\newcommand{\etalchar}[1]{$^{#1}$}
\begin{thebibliography}{vdAMSW09}

\bibitem[{Ant}12]{ant}
{Ant Project Contributors}.
\newblock {Apache Ant}.
\newblock [online], 2000--2012.
\newblock \url{http://ant.apache.org/}.

\bibitem[BBW{\etalchar{+}}08]{wsc08}
M.~Brian Blake, Steffen Bleul, Thomas Weise, Andreas Wombacher, Michael~C.
  Jaeger, William~K. Cheung, and Brian Miller.
\newblock 2008 {Web Services Challenge} in conjunction with {EEE'08} and
  {CEC'08}.
\newblock [online], July 2008.
\newblock \url{http://cec2008.cs.georgetown.edu/wsc08/interface_package.html},
  last viewed February 2011.

\bibitem[BT08]{standards-based-service-repo-2008}
Paul~A. Buhler and R.~W. Thomas.
\newblock Experiences building a standards-based service description
  repository.
\newblock In {\em CEC/EEE 2008}, pages 343--346, 2008.

\bibitem[CMt14]{marf-test-nn}
Ian Clement, Serguei~A. Mokhov, and {the MARF Research \& Development Group}.
\newblock {TestNN -- Testing Artificial Neural Network in MARF}.
\newblock Published electronically within the MARF project,
  \url{http://marf.sf.net}, 2002--2014.
\newblock Last viewed February 2010.

\bibitem[CT08]{soa-ws-data-management-2008}
Yinong Chen and Wei-Tek Tsai.
\newblock {\em Service-Oriented Computing and Web Data Management: From
  Principles to Development}.
\newblock Kendall/Hunt Publishing Company, first edition, 2008.
\newblock {ISBN}: 978-0-7575-7747-5. Online at
  \url{http://www.public.asu.edu/~ychen10/book/DSOSD.pdf}.

\bibitem[EAA{\etalchar{+}}04]{ibm-redbook-patterns-soa-ws-2004}
Mark Endrei, Jenny Ang, Ali Arsanjani, et~al.
\newblock {\em Patterns: Service-Oriented Architecture and Web Services}.
\newblock IBM, 2004.
\newblock IBM Red Book; online at
  \url{http://www.redbooks.ibm.com/abstracts/sg246303.html}.

\bibitem[FFK04]{fm-verification-bpel4ws-2004}
Jesus~Arias Fisteus, Luis~Sanchez Fernandez, and Carlos~Delgado Kloos.
\newblock Formal verification of bpel4ws business collaborations.
\newblock In K.~Bauknecht, M.~Bichler, and B.~Proll, editors, {\em EC-Web
  2004}, LNCS 3182, pages 76--85. Springer, 2004.

\bibitem[Fla97]{javanuttshell}
D.~Flanagan.
\newblock {\em Java in a Nutshell}.
\newblock O'Reily \& Associates, Inc., second edition, 1997.
\newblock {ISBN} 1-56592-262-X.

\bibitem[Gag11]{java-xml-parsing-dom-stringreader}
R\'{e}al Gagnon.
\newblock Parse an {XML} string using {DOM} and a {StringReader}.
\newblock [online], 1998--2011.
\newblock http://www.rgagnon.com/javadetails/java-0573.html, last viewed
  January 2011.

\bibitem[GNT04]{automated-planning-2004}
Malik Ghallab, Dana Nau, and Paolo Traverso.
\newblock {\em Automated Planning: Theory \& Practice}.
\newblock Morgan Kaufmann Publishers, May 2004.

\bibitem[Hen06]{rise-fall-corba-2006}
Michi Henning.
\newblock The rise and the fall of {CORBA}.
\newblock {\em ACM QUEUE}, pages 28--34, June 2006.
\newblock \url{https://queue.acm.org/detail.cfm?id=1142044}.

\bibitem[HMLW08]{ws-sim-map-forest-fire-2008}
Yosri Harzallah, Vincent Michel, Qi~Liu, and Gabriel Wainer.
\newblock Distributed simulation and web map mash-up for forest fire spread.
\newblock [online], 2008.

\bibitem[IBM{\etalchar{+}}07]{ws-bpel-11}
{IBM}, {BEA Systems}, {Microsoft}, {SAP AG}, and {Siebel Systems}.
\newblock {Business Process Execution Language for Web Services} version 1.1.
\newblock [online], IBM, February 2007.
\newblock
  \url{http://www.ibm.com/developerworks/library/specification/ws-bpel/}.

\bibitem[JY08]{ws-mash-up-home-library-2008}
Suresh Jeyaverasingam and Yuhong Yan.
\newblock Mash up home library system.
\newblock [online], 2008.

\bibitem[Koe07]{koenig-ws-bpel-2007}
Dieter Koenig.
\newblock Web services business process execution language ({WS-BPEL} 2.0): The
  standards landscape.
\newblock Presentation, IBM Software Group, 2007.

\bibitem[{Mav}12]{maven}
{Maven Project Contributors}.
\newblock {Apache Maven}.
\newblock [online], 2002--2012.
\newblock \url{http://maven.apache.org/}.

\bibitem[MC09]{vim}
Bram Moolenaar and {Contributors}.
\newblock {Vim} the editor -- {Vi Improved}.
\newblock [online], 2009.
\newblock \url{http://www.vim.org/}.

\bibitem[MJ08]{dmarf-web-services-cisse08}
Serguei~A. Mokhov and Rajagopalan Jayakumar.
\newblock {Distributed Modular Audio Recognition Framework} ({DMARF}) and its
  applications over web services.
\newblock In Tarek Sobh, Khaled Elleithy, and Ausif Mahmood, editors, {\em
  Proceedings of TeNe'08}, pages 417--422, University of Bridgeport, CT, USA,
  December 2008. Springer.
\newblock Printed in January 2010.

\bibitem[ML08]{wsc-gis-2008}
Sai Ma and Minruo Li.
\newblock Service composition for {GIS}.
\newblock [online], 2008.

\bibitem[Mok05]{mokhov-unix-commands}
Serguei~A. Mokhov.
\newblock {UNIX} commands, revision 1.4.
\newblock [online], 2003--2005.
\newblock
  \url{http://users.encs.concordia.ca/~mokhov/comp444/tutorials/unix-commands.pdf}.

\bibitem[Mok06]{dmarf06}
Serguei~A. Mokhov.
\newblock On design and implementation of distributed modular audio recognition
  framework: Requirements and specification design document.
\newblock [online], August 2006.
\newblock Project report, \url{http://arxiv.org/abs/0905.2459}, last viewed
  April 2012.

\bibitem[Mok09]{mokhov-dot-cshrc-example}
Serguei~A. Mokhov.
\newblock A \texttt{.cshrc} example, 2000--2009.
\newblock \url{http://users.encs.concordia.ca/~mokhov/comp346/.cshrc}.

\bibitem[Mor08]{java-http-post-get-requests}
Aviran Mordo.
\newblock Make {HTTP POST} or {GET} request from {Java}.
\newblock [online], January 2008.
\newblock
  \url{http://www.aviransplace.com/2008/01/08/make-http-post-or-get-request-from-java/},
  last viewed January 2011.

\bibitem[{Net}11a]{netbeans-kb-ws-jaxb}
{NetBeans Community}.
\newblock Binding {WSDL} to {Java} with {JAXB}.
\newblock [online], 2011.
\newblock \url{http://netbeans.org/kb/docs/websvc/jaxb.html}.

\bibitem[{Net}11b]{netbeans-kb-gs-axis}
{NetBeans Community}.
\newblock Creating {Apache} {Axis2} web services on {NetBeans IDE}.
\newblock [online], 2011.
\newblock \url{http://netbeans.org/kb/69/websvc/gs-axis.html}.

\bibitem[{Net}11c]{netbeans-kb-ws-jax-ws}
{NetBeans Community}.
\newblock Getting started with {JAX-WS} web services.
\newblock [online], 2011.
\newblock \url{http://netbeans.org/kb/docs/websvc/jax-ws.html}.

\bibitem[{Net}11d]{netbeans-kb-trails-web}
{NetBeans Community}.
\newblock Web services learning trail.
\newblock [online], 2011.
\newblock \url{http://netbeans.org/kb/trails/web.html}.

\bibitem[{Net}12]{netbeans-691}
{NetBeans Community}.
\newblock {NetBeans 6.9.1}.
\newblock [online], 2010--2012.
\newblock \url{http://netbeans.org/downloads/6.9.1/index.html}.

\bibitem[{Net}14]{netbeans}
{NetBeans Community}.
\newblock {NetBeans Integrated Development Environment}.
\newblock [online], 2004--2014.
\newblock \url{http://www.netbeans.org}.

\bibitem[{Net}15]{netbeans-kb-ws-rest}
{NetBeans Community}.
\newblock Getting started with {RESTful} web services.
\newblock [online], 2011--2015.
\newblock \url{http://netbeans.org/kb/docs/websvc/rest.html}.

\bibitem[NKL08]{type-aware-wsc-bool-sat-2008}
Wonhong Nam, Hyunyoung Kil, and Dongwon Lee.
\newblock Type-aware web service composition using boolean satisfiability
  solver.
\newblock In {\em CEC/EEE 2008}, pages 331--334, 2008.

\bibitem[{OAS}07]{ws-bpel-20}
{OASIS Web Services Business Process Execution Language (WSBPEL) TC}.
\newblock {Web Services Business Process Execution Language} version 2.0.
\newblock [online], Oasis, April 2007.
\newblock OASIS Standard,
  \url{http://docs.oasis-open.org/wsbpel/2.0/OS/wsbpel-v2.0-OS.html}.

\bibitem[OL09]{wsben-2009}
Seog-Chan Oh and Dongwon Lee.
\newblock {WSBen}: A web services discovery and composition benchmark toolkit.
\newblock {\em Int. J. Web Service Res.}, 6(1):1--19, 2009.

\bibitem[OLK05]{comparative-ill-ai-wsc-2005}
Seog-Chan Oh, Dongwon Lee, and Soundar R.~T. Kumara.
\newblock A comparative illustration of {AI} planning-based web services
  composition.
\newblock {\em ACM SIGecom Exchanges}, 5(5):1--10, December 2005.

\bibitem[OLK07]{wspr-scalable-wsc-algo-2007}
Seog-Chan Oh, Dongwon Lee, and Soundar R.~T. Kumara.
\newblock Web service planner (wspr): An effective and scalable web service
  composition algorithm.
\newblock {\em Int. J. Web Service Res.}, 4(1):1--22, 2007.

\bibitem[{Ope}09]{bpelse}
{OpenESB Contributors}.
\newblock {BPEL} service engine.
\newblock [online], 2009.
\newblock \url{https://open-esb.dev.java.net/BPELSE.html}.

\bibitem[OvdAB{\etalchar{+}}05]{fm-semantics-ctrl-flow-ws-bpel-2005}
C.~Ouyang, W.~van~der Aalst, S.~Breutel, M.~Dumas, A.~ter Hofstede, and
  H.~Verbeek.
\newblock Formal semantics and analysis of control flow in {WS-BPEL}.
\newblock Technical Report BPM-05-13, BPM Center Report, 2005.
\newblock \url{BPMcenter.org}.

\bibitem[PBB{\etalchar{+}}04]{planning-monitoring-wsc-2004}
Marco Pistore, Fabio Barbon, Piergiorgio Bertoli, Dmitry Shaparau, and Paolo
  Traverso.
\newblock Planning and monitoring web service composition.
\newblock In {\em AIMSA 2004}, pages 106--115, 2004.

\bibitem[Pee05]{wsc-ai-planning-survey}
Joachim Peer.
\newblock Web service composition as {AI} planning, a survey.
\newblock [online], 2005.
\newblock Second revised version,
  \url{http://citeseerx.ist.psu.edu/viewdoc/summary?doi=10.1.1.85.9119}.

\bibitem[SBS04]{reasoning-ws-process-algebra-2004}
G.~Sala\"{u}n, L.~Bordeaux, and M.~Schaerf.
\newblock Describing and reasoning on web services using process algebra.
\newblock In {\em Proceedings of 2nd International Conference on Web Services
  (ICWS04)}, 2004.

\bibitem[Seb10]{programming-www-2010}
Robert~W. Sebesta.
\newblock {\em Programming the {World Wide Web} 2010}.
\newblock Addison-Wesley, 6 edition, 2010.
\newblock {ISBN}: 978-0-13-213081-3.

\bibitem[SH05]{soa-semantics-processes-agents-2005}
Munindar~P. Singh and Michael~N. Huhns.
\newblock {\em Service-Oriented Computing: Semantics, Processes, Agents}.
\newblock John Wiley \& Sons, Ltd, West Sussex, England, 2005.

\bibitem[{Sun}05a]{servlets}
{Sun Microsystems, Inc.}
\newblock Java servlet technology.
\newblock [online], 1994--2005.
\newblock \url{http://java.sun.com/products/servlets}.

\bibitem[{Sun}05b]{jsp}
{Sun Microsystems, Inc.}
\newblock {JavaServer} pages technology.
\newblock [online], 2001--2005.
\newblock \url{http://java.sun.com/products/jsp/}.

\bibitem[{Sun}06a]{java-webservices-jaxb}
{Sun Microsystems, Inc.}
\newblock The {Java} web services tutorial: Binding {XML} schemas.
\newblock [online], February 2006.
\newblock
  \url{http://download.oracle.com/docs/cd/E17802_01/webservices/webservices/docs/2.0/tutorial/doc/JAXBWorks4.html}.

\bibitem[{Sun}06b]{java-webservices}
{Sun Microsystems, Inc.}
\newblock The {Java} web services tutorial (for {Java Web Services Developer's
  Pack}, v2.0).
\newblock [online], February 2006.
\newblock
  \url{http://download.oracle.com/docs/cd/E17802_01/webservices/webservices/docs/2.0/tutorial/doc/}.

\bibitem[{Sun}09a]{glassfish}
{Sun Microsystems, Inc.}
\newblock {Sun GlassFish}: Open web application platform.
\newblock [online], 1994--2009.
\newblock \url{http://www.sun.com/glassfish}.

\bibitem[{Sun}09b]{netbeans-651}
{Sun Microsystems, Inc.}
\newblock {NetBeans 6.5.1}.
\newblock [online], July 2004--2009.
\newblock \url{http://netbeans.org/downloads/6.5.1/index.html}.

\bibitem[{Sun}10]{netbeans-671}
{Sun Microsystems, Inc.}
\newblock {NetBeans 6.7.1}.
\newblock [online], 2009--2010.
\newblock \url{http://netbeans.org/downloads/6.7.1/index.html}.

\bibitem[{The}14]{marf}
{The MARF Research and Development Group}.
\newblock {The Modular Audio Recognition Framework and its Applications}.
\newblock [online], 2002--2014.
\newblock \url{http://marf.sf.net} and \url{http://arxiv.org/abs/0905.1235},
  last viewed May 2015.

\bibitem[vdAMSW09]{service-interaction-2009}
Wil M.~P. van~der Aalst, Arjan~J. Mooij, Christian Stahl, and Karsten Wolf.
\newblock Service interaction: Patterns, formalization, and analysis.
\newblock In {\em SFM 2009}, LNCS 5569, pages 42--88. Springer, June 2009.

\bibitem[Vir04]{fm-orchestration-2004}
M.~Viroli.
\newblock Towards a formal foundation to orchestration languages.
\newblock {\em Electronic Notes in Theoretical Computer Science}, 105:51--71,
  2004.

\bibitem[VL08]{ws-soap-routing-2008}
Wiwat Vatanawood and Weerakiat Limpichotipong.
\newblock A web service request routing system through co-agreement {SOAP}
  routers.
\newblock {\em International Journal of Web Services Practices}, 3(4):136--139,
  2008.

\bibitem[{Wik}09]{wiki:bpel}
{Wikipedia}.
\newblock {Business Process Execution Language (BPEL)} --- {Wikipedia}{,} the
  free encyclopedia.
\newblock [Online; accessed 14-July-2009], 2009.
\newblock
  \url{http://en.wikipedia.org/w/index.php?title=Business_Process_Execution_Language&oldid=302021294}.

\bibitem[YABCZ10]{compat-reparation-ws-processes-2010}
Yuhong Yan, Ali A\"{\i}t-Bachir, Min Chen, and Kai Zhang.
\newblock Compatibility and reparation of web service processes.
\newblock In {\em ICWS}, pages 634--637, 2010.

\bibitem[Yan08]{dl-fm-wsp-modeling-2008}
Yuhong Yan.
\newblock Description language and formal methods for web service process
  modeling.
\newblock In {\em Business Process Management: Concepts, Technologies and
  Applications}, volume Advances in Management Information Systems. M.E. Sharpe
  Inc., 2008.

\bibitem[Yan11a]{soen-691a-487-winter2011-xml}
Yuhong Yan.
\newblock Service computing: Basics of {XML}.
\newblock [online], 2011.
\newblock Chapter 3.

\bibitem[Yan11b]{soen-691a-487-winter2011-bpel}
Yuhong Yan.
\newblock Service computing: {BPEL: Business Process Execution Language} for
  web services.
\newblock [online], 2011.

\bibitem[Yan11c]{soen-691a-487-winter2011-ws-programming}
Yuhong Yan.
\newblock Service computing: Coding with {SOAP} web services.
\newblock [online], 2011.

\bibitem[Yan11d]{soen-691a-487-winter2011}
Yuhong Yan.
\newblock Service computing: Course notes ({SOEN691A} and {SOEN487}).
\newblock [online], 2011.

\bibitem[Yan11e]{soen-691a-487-winter2011-rest}
Yuhong Yan.
\newblock Service computing: Develop {RESTful} services.
\newblock [online], 2011.

\bibitem[Yan11f]{soen-691a-487-winter2011-distrib-computing}
Yuhong Yan.
\newblock Service computing: Distributed computing infrastructure.
\newblock [online], 2011.
\newblock Chapter 2.

\bibitem[Yan11g]{soen-691a-487-winter2011-intro}
Yuhong Yan.
\newblock Service computing: Introduction to web services and service
  computing.
\newblock [online], 2011.
\newblock Chapter 1.

\bibitem[Yan11h]{soen-691a-487-winter2011-uddi}
Yuhong Yan.
\newblock Service computing: Registering and discovering web services.
\newblock [online], 2011.
\newblock Chapter 6.

\bibitem[Yan11i]{soen-691a-487-winter2011-service}
Yuhong Yan.
\newblock Service computing: Service, service systems and services innovation.
\newblock [online], 2011.

\bibitem[Yan11j]{soen-691a-487-winter2011-soap}
Yuhong Yan.
\newblock Service computing: {SOAP}: Simple object access protocol.
\newblock [online], 2011.
\newblock Chapter 4.

\bibitem[Yan11k]{soen-691a-487-winter2011-wsdl}
Yuhong Yan.
\newblock Service computing: {WSDL}: Web service definition language.
\newblock [online], 2011.
\newblock Chapter 5.

\bibitem[YBWM08]{service-science-and-soss-2008}
Yuhong Yan, Juergen Bode, and Jr. William~McIver.
\newblock Between service science and service-oriented software systems.
\newblock In {\em ICWS 2008}, 2008.

\bibitem[YDPC09]{model-faults-in-ws-processes-2009}
Yuhong Yan, Philippe Dague, Yannick Pencole, and Marie-Odile Cordier.
\newblock A model-based approach for diagnosing faults in web service
  processes.
\newblock {\em International Journal on Web Service Research}, 6(1):87--110,
  2009.

\bibitem[YKLO08]{wsc-framework-int-programming-2008}
Jung-Woon Yoo, Soundar R.~T. Kumara, Dongwon Lee, and Seog-Chan Oh.
\newblock A web service composition framework using integer programming with
  non-functional objectives and constraints.
\newblock In {\em CEC/EEE 2008}, pages 347--350, 2008.

\bibitem[YLRD09]{ws-enabled-lab-2009}
Yuhong Yan, Yong Liang, Abhijeet Roy, and Xinge Du.
\newblock Web service enabled online laboratory.
\newblock {\em International Journal on Web Service Research}, 2009.

\bibitem[YPZ10]{repairing-service-composition-2010}
Yuhong Yan, Pascal Poizat, and Ludeng Zhao.
\newblock Repairing service compositions in a changing world.
\newblock In Roger Lee, Olga Ormandjieva, Alain Abran, and Constantinos
  Constantinides, editors, {\em Proceedings of SERA 2010 (selected papers)},
  volume 296 of {\em Studies in Computational Intelligence}, pages 17--36.
  Springer Berlin Heidelberg, 2010.

\bibitem[YXG08]{auto-wsc-and-or-graph-2008}
Yixin Yan, Bin Xu, and Zhifeng Gu.
\newblock Automatic service composition using {AND/OR} graph.
\newblock In {\em CEC/EEE 2008}, pages 335--338, 2008.

\bibitem[YZ08]{plan-graph-algo-semantic-wsc-2008}
Yuhong Yan and Xianrong Zheng.
\newblock A planning graph based algorithm for semantic web service
  composition.
\newblock In {\em CEC/EEE 2008}, pages 339--342, 2008.

\end{thebibliography}

\printindex

\end{document}